\documentclass[fleqn,10pt]{wlscirep}
\usepackage[utf8]{inputenc}
\usepackage[T1]{fontenc}

\title{Charge-optimized many-body interaction potential for AlN revisited to explore plasma-surface interactions}

\author[1]{Tobias Gergs}
\author[1]{Thomas Mussenbrock}
\author[2,3,*]{Jan Trieschmann}

\affil[1]{Chair of Applied Electrodynamics and Plasma Technology, Department of Electrical Engineering and Information Science, Ruhr University Bochum, 44780 Bochum, Germany}
\affil[2]{Theoretical Electrical Engineering, Department of Electrical and Information Engineering, Kiel University, Kaiserstraße 2, 24143 Kiel, Germany}
\affil[3]{Kiel Nano, Surface and Interface Science KiNSIS, Kiel University, Christian-Albrechts-Platz 4, 24118 Kiel, Germany}

\affil[*]{jt@tf.uni-kiel.de}

\begin{abstract}
Plasma-surface interactions during AlN thin film sputter deposition could be studied by means of reactive molecular dynamics (RMD) methods. This requires an interaction potential that describes all species as well as wall interactions (e.g., particle emission, damage formation) appropriately. However, previous works focused on the establishment of AlN bulk potentials. Although for the third-generation charge-optimized many-body (COMB3) potential at least a single reference surface was taken into account, surface interactions are subject to limited reliability only. The demand for a revised COMB3 AlN potential is met in two steps: First, the Ziegler-Biersack-Littmark potential is tapered and the variable charge model QTE$^+$ is implemented to account for high-energy collisions and distant charge transport, respectively. Second, the underlying parameterization is reworked by applying a self-adaptive evolution strategy implemented in the GARFfield software. Four wurtzite, three zinc blende and three rock salt surfaces are considered. An example study on the ion bombardment induced particle emission and point defect formation reveals that the revised COMB3 AlN potential is appropriate for the accurate investigation of plasma-surface interactions by means of RMD simulations.
\end{abstract}
\begin{document}

\flushbottom
\maketitle

\thispagestyle{empty}

\section*{Introduction}

The powering of micro-electro-mechanical systems and wireless sensors can be realized by energy harvesting (scavenging), overcoming the limitations of batteries (e.g., runtime, maintenance costs). Absorbed ambient vibrations can be translated to mechanical stresses of a piezoelectric material. The induced strain causes a polarization and, hence, voltage drop across the material. Hexagonal aluminium nitride (AlN) is argued to be a promising material system for such an application due its high chemical, thermal, and mechanical stability \cite{iqbal_reactive_2018,elfrink_vibration_2009,batra_piezoelectric_2016,beeby_energy_2006}. Its properties make it also a widely used subsystem for hard coatings and protective wear applications (e.g., transition metal aluminium nitride, transition metal aluminium oxynitride) \cite{gibson_quantum_2018}. For either application (i.e., energy harvesting, hard coatings) sputter deposition is exploited to fabricate the thin films. A plasma is ignited between two opposing electrodes by applying a voltage of up to 1 kV. Background gas atoms (e.g., Ar, N, N$_2$) are ionized and accelerated toward the (Al) target electrode. The ion bombardment spawns a collision cascade within the material which eventually leads to the emission of surface atoms. After transport through the gas phase these atoms may impinge onto the opposing substrate surface. The simultaneous flux of neutral background gas atoms and (less energetic) ion bombardment on the substrate surface together result in sputter deposited (AlN) thin film growth. The film properties (e.g., stress, defect structure) intrinsically determine the performance during application. Hence, it is fundamental to truly understand and possibly design the defect formation during deposition.

The required information can accessed by reactive molecular dynamics (RMD) simulations of AlN growth or sputtering processes. However, the fidelity of RMD studies inherently relies on the accuracy of the applied interaction potential. A variety of potentials (e.g., Vashishta, Tersoff) has been proposed for the investigation of AlN bulk systems and processes (e.g., pressure induced phase changes) \cite{vashishta_interaction_2011,tungare_tersoff-based_2011}. A detailed comparison can be found elsewhere \cite{xiang_comparison_2017}. Surfaces have only been taken into account for setting up the third-generation charge-optimized many-body (COMB3) AlN potential, which is also able to describe Al and N$_2$ \cite{choudhary_dynamical_2016,choudhary_charge_2014}. However, the underlying database consisits of only a single surface energy and four point defect formation energies. This shortcoming may be of little relevance for the intended application (i.e., nanostructures, heterogeneous interfaces), but the lack of relevant reference data may cause issues for growth or sputtering simulations \cite{choudhary_dynamical_2016}. Notably, the published COMB3 TiN, TiO$_2$ and Al$_2$O$_3$ potentials may provide a basis for developing COMB3 potentials for hard coatings such as TiAlN or even TiAlON in the future. \cite{choudhary_dynamical_2016,cheng_charge_2014,choudhary_charge_2015,liang_classical_2013}.

In this work, the demand for an AlN interaction potential that allows for the investigation of ion bombardment induced defect formation is met by revising the COMB3 AlN potential. First, the applied methods are detailed. Second, the results on the performance of the revised potential and an example study of Ar$^+$, Al$^+$, N$^+$ and N$_2^+$ ions bombarding an AlN(0001) surface are gathered. Finally, conclusions are drawn.

\section*{Methods}

In the following, a summary of the implemented extensions of the COMB3 potential for the AlN system is provided. A detailed description of the general COMB3 formalism and its overall capabilities can be found elsewhere \cite{liang_classical_2013}. This is followed by a discussion of the assembled reference data used to revise the COMB3 AlN potential, and the genetic algorithm based reactive force field optimizer (GARFfield) and its extensions \cite{jaramillo-botero_general_2014}. The molecular dynamics methods used to conduct an example study with the revised COMB3 potential are outlined. The surface interaction of isolated ions (i.e., Al$^+$, N$^+$, Ar$^+$, N$_2^+$) impinging onto wurtzite AlN(0001) is analyzed.
In the following, the original COMB3 potential parameters for AlN due to Choudhary are denoted by COMB3$_\mathrm{QEq}^{2016}$ \cite{choudhary_dynamical_2016}. The publicly released COMB3 AlN parameters (which notably differ) are referred to by COMB3$_\mathrm{QEq}^{2016^*}$ \cite{COMB3AlNRelease}. The revised potential parameters presented in the following are denoted by COMB3$_{\mathrm{QTE}^+}^{2022}$.

\subsection*{Interaction potential}

In the frame of COMB3, a system that consists of two atoms of one kind, is described by short range and van der Waals interactions. The latter is often neglected due to its relatively weak contribution. When two atoms $i$ and $j$ differ in their elemental type, charge is transferred from on to the other. The charge of each atom consists of the particular nucleus charge $Z$, which is modeled as a point charge, and $q-Z$, which is spatially described by a 1s Slater type orbtial (STO). If required, point dipoles can be taken into account to model the system's electrostatic interaction more accurately. The charge transport takes place on the electronic time scale. As a consequence of the mass ratio, the nucleus can be thought of as immobile during the electron dynamics. In combination with COMB3, the corresponding equilibration process is typically modeled by the charge equilibration (QEq) method in conjunction with an extended Lagrangian formalism \cite{liang_classical_2013,rappe_charge_1991}. However, QEq fails to describe the charge transport during most relevant surface interactions (e.g., adsorption, desorption, deposition, sputtering). The application of the charge transfer equilibration (QTE$^+$) in favor of the QEq model is thus recommended for studies that target surface interactions \cite{gergs_generalized_2021}. The 1s STOs are required to enable a matching charge transfer per unit time. The corresponding electrostatic long range interactions are described in detail elsewhere \cite{liang_classical_2013,rappe_charge_1991}. 

For a binary system, the initially mentioned short range interactions consist of a repulsive as well as an attractive pair function $V^\mathrm{A}$ and $V^\mathrm{R}$, respectively \cite{liang_classical_2013,rappe_charge_1991}:
\begin{subequations}
\begin{align}
V_{ij}^\mathrm{R} &= A_{ij}\exp\left(-\lambda_{ij}r_{ij}+\lambda_{i}^*\frac{D_i(q_i)+D_j(q_j)}{2}\right)
\label{eq:VA}\\
V_{ij}^\mathrm{A} &= B_{ij}^*(q_i,q_j)\exp\left(\alpha_i^*\frac{D_i(q_i)+D_j(q_j)}{2}\right)\sum_{n=1}^3 B_{ij}^n\exp\left(-\alpha_{ij}^nr_{ij}\right)
\label{eq:VR}
\end{align}
\end{subequations}
$r_{ij}$ is the distance between atom $i$ and atom $j$. A detailed description of the per element parameter $\lambda_i^*,~D_i(q_i),~D_j(q_j),~\alpha_i^*$ and $B_ij^*(q_i,q_j)$ can be found in \cite{liang_classical_2013,rappe_charge_1991}. Only the pair parameters $A_{ij},~\lambda_{ij},~B_{ij}^0$ and $\alpha_{ij}^0$ are modified for Al-N interactions in this work ($B_{ij}^1=B_{ij}^2=\alpha_{ij}^1=\alpha_{ij}^2=0$).

The attractive term $V^\mathrm{A}$ is scaled with the average bond order of both atoms $\frac{b_{ij+b_{ji}}}{2}$ which allows for a more flexible and environment dependent potential \cite{liang_classical_2013,rappe_charge_1991}. The bond order $b_{ij}$ in a carbon exclusive system (i.e., AlN) is defined as follows
\begin{equation}
b_{ij}=\left(1+\sum_{k=1,k\neq\{i,j\}}^N f_\mathrm{c}(r_{ik})\zeta_{ij}(r_{ij},r_{ik})\sum_{n=0}^6 b_{n,ijk}\cos(\theta_{ijk})^n+P_{ij}(\Omega_i)\right)^{-\frac{1}{2}}
\label{eq:bond_order}
\end{equation}
$f_\text{c}$ is the Tersoff cutoff function. $\zeta_{ij}=\exp(\omega_{ij}(r_{ij}-r_{ik}))$ penalizes bond asymmetries. The parameter $\omega_{ij}$ is used to scale its impact \cite{liang_classical_2013,rappe_charge_1991}. It is adjusted in this work for N-N, Al-N and N-Al pairs. The coefficients $b_{n,ijk}$ of a sixth order polynomial are used to introduce a dependency on the bond angle $\theta_{ijk}$. The $i$-th atom is the centering atom. These parameters are three dimensional, whereas in the original COMB3 formalism they are intended to be two dimensional (i.e., $b_{n,ij}$) \cite{liang_classical_2013,rappe_charge_1991}. They are fitted in this work for Al-Al-N, Al-N-Al, Al-N-N, N-Al-Al, N-Al-N and N-N-Al. The effect of the coordination number $\Omega_i$ is added by $P_{ij}(\Omega_i)=c_{0,ij}\Omega_i+c_{1,ij}\exp(c_{2,ij}\Omega_i)+c_{3,ij}$ for Al-N and N-Al. 

Legendre polynomials up to the sixth order are added as a function of $\cos(\theta_{ijk})$ to the potential energy to ease the differentiation between face-centred cubic (fcc) and hexagonal close-packed (hcp) phases \cite{liang_classical_2013,rappe_charge_1991}. The corresponding coefficients $K_{ijk}^\mathrm{LP1}$, $K_{ijk}^\mathrm{LP2}$, $K_{ijk}^\mathrm{LP3}$, $K_{ijk}^\mathrm{LP4}$, $K_{ijk}^\mathrm{LP5}$ and $K_{ijk}^\mathrm{LP6}$ are adjusted for Al-Al-N (Al-N-Al), Al-N-N, N-Al-Al and N-Al-N (N-N-Al). A symmetry is imposed that demands $K_{ijk}^\mathrm{LP1}=K_{ikj}^\mathrm{LP1}$.

High-energy collisions during the ion bombardment are accounted for by implementing a combined (tapered) version of the revised COMB3$_\mathrm{QTE^+}^{2022}$ with the Ziegler-Biersack-Littmark (ZBL) potential 
in the Large-scale Atomic/Molecular Massively Parallel Simulator (LAMMPS) \cite{ziegler_stopping_1985,thompson_lammps_2022,plimpton_fast_1995,gergs_molecular_2022}. For $r_\mathrm{th,2} < r_{ij}$, the interaction is purely described by the COMB3$_\mathrm{QTE^+}^{2022}$ potential. In between $r_\mathrm{th,1}<r_{ij}\leq r_\mathrm{th,2}$, the interactions are determined by the tapered interaction potential.
The contribution of the ZBL potential is gradually increased with decreasing distance by scaling with the Tersoff cutoff function. For $r_{ij} \leq r_\mathrm{th,1}$, the interaction is purely described by the ZBL potential. The values for $r_\mathrm{th,1}$ and $r_\mathrm{th,2}$ for the considered element pairs are listed in Table~\ref{table: tapering_thres}. Bond-order definitions and calculations are not changed by the implemented tapering. Hence, two atoms with $r_\mathrm{th,1}<r_{ij}\leq r_\mathrm{th,2}$ contribute to each other's bond-order. 

\begin{table}[ht]
\centering
\begin{tabular}{|l|c|c|c|c|}
\hline
& Al-Al & Al-N & N-N & Ar-Ar\\
\hline
$r_\mathrm{th,1}$ (\r A) & 0.95 & 0.7 & 0.425 & 2.25 \\
\hline
$r_\mathrm{th,2}$ (\r A) & 1.9 & 1.4  & 0.85 & 3.05 \\
\hline
\end{tabular}
\caption{\label{table: tapering_thres}Threshold values $r_\mathrm{th,1}$ and $r_\mathrm{th,2}$ for tapering COMB3$_\mathrm{QTE^+}^{2022}$/ZBL and LJ/ZBL.}
\end{table}

A similar tapering of the Lennard-Jones (LJ) potential (with an outer cutoff radius of 8.5 \r A) with the ZBL potential is used for Ar-Ar interactions \cite{ziegler_stopping_1985}. The values for $r_\mathrm{th,1}$ and $r_\mathrm{th,2}$ are also listed in Table~\ref{table: tapering_thres}. The depths of the LJ potential well and zero-crossing distance are 0.0104 and 3.4 \r A, respectively. The Al-Ar and N-Ar interactions are described purely by the ZBL potential (with an outer cutoff radius of 8.0 \r A).

The implementations are verified by comparing the integrated force with the corresponding potential energy and the tapered with the non-tapered potentials for a relevant set of cases (see Supplementary Methods). 

\subsection*{Reference data}

The assembled reference data set used to revise the COMB3$^{2016^*}$ AlN potential is listed in the following:

\begin{itemize}
    \item 28 Al$_x$N$_y$ clusters with up to eight atoms (i.e., $x\leq6$, $y\leq4$) (resembling structures on top a surface or during ion bombardment induced surface reconstructions). They are described by the atoms' geometry (i.e., bond lengths, bond angles), and (relative) binding energies.
    \item Four ideal bulk lattice structures (i.e., wurtzite (WZ), zinc blende (ZnB), rock salt (NaCl), CsCl). They are described by 12 cell parameters, eight elastic constants (i.e., five for the wurtzite phase, three for the zinc blende phase), three heats of formations, and three heats of reactions.
    \item Five (seven) point defect structures for the wurtzite (zinc blende) phase. They are described by defect formation energies (N rich conditions are assumed).
    \item Four wurtzite, three zinc blende, and three rock salt surfaces are taken into account. They are described by their surface energies.
    \item 150 bond lengths and angles. They describe the atom geometries introduced above.
\end{itemize}


\subsection*{GARFfield extensions}

The software GARFfield developed by Jaramillo Botero et al. is a reactive force field optimizer that is described in detail in \cite{jaramillo-botero_general_2014}. In the following a brief summary focusing on implemented extensions, minor revisions, and the herein applied features is presented.

Multiple interaction potentials (e.g., ReaxFF, eFF, COMB) are included in GARFfield. The COMB potential and the QEq method are replaced in this work with the COMB3 potential and the QTE$^+$ method, respectively. The reading (writing) of the interaction potential files are revised accordingly. The COMB3 AlN parameters (genes) to be fitted are listed in the preceding section. This list is referred to as a chromosome with length $n$ in the frame of a genetic algorithm (GA) as it is applied within GARFfield \cite{jaramillo-botero_general_2014}. An initial population with diverse chromosomes and genes is established by sampling randomly in a given interval around the COMB3$_\mathrm{QEq}^{2016^*}$ AlN parameters. The lower and upper bound of the $i$-th gene are $x_{\mathrm{lo},i}$ and $x_{\mathrm{hi},i}$, respectively. The fitness of each chromosome is evaluated by performing molecular statics calculations of the atom configurations mentioned in the preceding section, with the LAMMPS library (including COMB3$_\mathrm{QTE^+}$) \cite{thompson_lammps_2022,plimpton_fast_1995}. A conjugate gradient descent algorithm is used to relax the individual structures. The time step for the QTE$^+$ fictitious charge motions is chosen as $10^{-2}$ fs. The tolerance for the maximum charge force and force on any atom is 10$^{-2}$ V and 1 eV/\r A, respectively. These evaluations are handled in a parallel manner by utilizing a modified version of PGAPack in GARFfield \cite{levine_users_1996,jaramillo-botero_general_2014}. A master-slave architecture is established, in which each processor performs the molecular statics calculations in serial for each individual out of the population.

The list of available reference properties (e.g., cell parameters, formation energies) is enlarged by the implementation of elastic constant and surface energy calculations. Moreover, the order of the program execution is altered to enable more consistent evaluations. First, the relaxed cell parameters are determined. Second, these cell parameters are then used to scale other atom configurations (e.g., point defect structures). 

Deviation between the fitted and the reference properties are quantified by the weighted mean percentage error (WMPE). The sectional weights for the cell parameters, atom geometries, elastic constants, and energies are 100, 3, 1, 12, respectively. The weights for each entry in the cell parameters, atom geometries, and elastic constants are chosen to even out the impact of each atom structure. The weights for the energies are chosen to even out the impact of i) wurzite AlN heat of formation; ii) relative zinc blende, rock salt and CsCl heats of formation (heats of reaction); iii) binding energies of AlN clusters; iv) wurtzite  point defect formation energies; v) zinc blende point defect formation energies; vi) surface energies. However, the WMPE is still ill defined for a small number of reference samples. This issue is resolved by defining a limit for the denominator value of 1 eV, 0.1 GPa and 0.1 eV/\r A$^2$ for binding (formation) energies, elastic constants, and surface energies, respectively.

As a metric for the fitness of the evolution, the WMPE is used to rank and select parents for reproduction and, hence, creation of the next population. The new chromosomes are obtained by applying two-point-crossover recombination to the parents' chromosomes. Genes are then randomly selected per chromosome for mutation. The probability is commonly chosen as the inverse of the chromosome length. Thus, mutating only one gene is the most probable event \cite{rozenberg_handbook_2012}. The value $x_i$ of the $i$-th gene is updated by $x_i\rightarrow\sigma\mathcal{N}_i(0,1)x_i$, whereas $\sigma$ and $\mathcal{N}_i(0,1)$ are the isotropic mutation strength and the standard normal distribution, respectively.

On the one hand, two-point-crossover recombination is found to hinder substantial evolutionary progress. This is reasoned by the specific linkage and setup of the COMB3 potential parameters. On the other hand, a GA without crossover recombination is considered to be just an inefficient evolution strategy (ES). Hence, we implemented an anisotropic self-adaptive evolution strategy with intermediate recombination $(\mu/\mu_\mathrm{I}/\lambda)$-$\sigma$SA-ES in GARFfield \cite{jaramillo-botero_general_2014}. $\mu$ and $\lambda$ are the number of parents and population size, respectively. Anisotropic self-adaptation means that the formerly constant mutation strength (step size) $\sigma$ is replaced with an evolving mutation strength (step size) $\sigma_i$ for each gene $i$. The list of these individual mutation strengths (step sizes) is appended to the chromosome. Hence, its length $n$ is doubled. Intermediate recombination means that the offspring population's chromosomes are at first set equal to the centroid of the parents' chromosomes. Intermediate recombination is implemented in the modified PGAPack software \cite{levine_users_1996}. All genes are then mutated. First, each mutation strength $\sigma_i$ is updated following
$\sigma_i\rightarrow \sigma_i \exp(\tau^\prime \mathcal{N}(0,1) + \tau \mathcal{N}_i(0,1))$. Random realizations are sampled once from $\tau^\prime \mathcal{N}(0,1)$ and $n$ times from $\tau \mathcal{N}_i(0,1)$ per generation to provide the flexibility for changes on the global and individual scale, respectively. The factors $\tau^\prime$ and $\tau$ are defined by $\tau^\prime=\sqrt{2\sqrt{n}}^{-1}$ and $\tau=\sqrt{2n}^{-1}$, respectively \cite{back_basic_1994,schwefel_numerische_1977,rozenberg_handbook_2012}. Second, each gene $x_i$ is updated by $x_i\rightarrow\sigma_i\mathcal{N}_i(0,1)(x_{\mathrm{hi},i}-x_{\mathrm{lo},i})/2+x_i$. Boundaries for each gene demanded by the COMB3 potential formalism are enforced (e.g., no square root of negative numbers). The lower (upper) bound of the initialization interval for the $i$-th gene [$x_{\mathrm{lo},i}$,$x_{\mathrm{hi},i}$] is chosen to be its initial value minus (plus) 20 \% when the gene directly effects either the bond-order or Legendre polynomial computations. When the gene effects only the attractive or repulsive pair potential, this percentage value is lowered by a factor of 10 (i.e., $\pm$ 2 \%). 

Overall the modified GARFfield implementation allows for a consistent and continuous transition from a GA to an ES, making it a more diverse evolutionary algorithm (EA).


The (1,79)-$\sigma$SA-ES implemented in GARFfield is applied to revise the COMB3$^{2016^*}$ AlN potential. Ten iterations each consisting of up to 200 generations have been performed. For the final run, the range of the initialization interval is reduced by a factor of four (i.e., $\pm$ 5 \%, $\pm$ 0.5 \%).

\subsection*{Exemplary study setup: Ion impingement on AlN(0001)}

As an exemplary study, the isolated impingement of ions -- in the present case Al$^+$, N$^+$, Ar$^+$, and N$_2^+$ -- onto the ideal wurtzite Al-AlN(0001) surface is investigated. The bombardment-induced mean damage formation and the flux emission are considered. Ion energies $E_\mathrm{ion}$ range from 10 eV to 100 eV in 10 eV steps. This range is chosen on the basis of experimentally measured mean ion energies impinging onto the substrate during the reactive radio frequency sputtering of AlN \cite{ries_ion_2019}. For each ion species and energy, the simulation is repeated 50 times, whereas the ions' lateral position is sampled from a uniform distribution and scaled with the surface dimensions. The damage formation is quantified by means of the induced point defect populations (i.e., $\rho_\mathrm{v_{Al}}$, $\rho_\mathrm{Al_N}$, $\rho_\mathrm{Al_i}$, $\rho_\mathrm{v_N}$, $\rho_\mathrm{N_{Al}}$, $\rho_\mathrm{(\text{N-N})_N}$, $\rho_\mathrm{Ar_{Al}}$, $\rho_\mathrm{Ar_N}$, $\rho_\mathrm{Ar_i}$).

LAMMPS is used for all simulations outlined in this subsection \cite{thompson_lammps_2022,plimpton_fast_1995}. First, a bulk $7\times4\times7$ wurtzite AlN system with lattice constants $a=3.154$ and $c=1.6a$ is constructed. It is shifted and cleaved in [001] direction to obtain the Al terminated Al-AlN(0001) surface. Bottom layers of the surface slab are excluded from time integration up to a height $z_\mathrm{th}$ to anchor it in the simulation domain, reduce the effect of the N-AlN(000$\overline{1}$) surface on the damage formation and reduce the computational cost. The variable $z_\mathrm{th}$ is defined through $E_\mathrm{ion}$ as follows $z_\mathrm{th}=4c-3c({\sqrt{E_\mathrm{ion}}-\sqrt{10~\mathrm{eV}}})/({\sqrt{100~\mathrm{eV}}-\sqrt{10~\mathrm{eV}}})$. Four ($E_\mathrm{ion}$=10 eV) to one ($E_\mathrm{ion}$=100 eV) AlN layers are kept immobile. The atom charge distribution is calculated on each time step by applying the QTE$^+$ method again with a time step of $10^{-2}$ fs and a tolerance for the maximum charge force of $10^{-2}$ V. The fitted STOs $\zeta^\mathrm{ov}$ for Al and N are 0.668 \r A$^{-1}$ and 1.239 \r A$^{-1}$, respectively \cite{gergs_generalized_2021}. 

This exemplary case study is performed twice to assess the necessity for equilibrating the surface prior to a particle impact: 
a) The mobile surface atoms are equilibrated at 300 K by applying a Berendensen thermostat to two AlN layers with $z_\mathrm{th}<z<z_\mathrm{th}+2c$ \cite{berendsen_molecular_1984}. A damping constant and time step of 100 fs and 0.25 fs is used, respectively. The simulations are run for 100 ps.
b) To obtain a surface relaxed initially at 0 K, atom positions are adjusted by applying a conjugate gradient minimization until the force on any atom is below 1 eV/\r A. For the latter case, a lower bound for the final temperature can be predicted
\begin{equation}
T_\mathrm{pred}=\frac{E_\mathrm{ion}}{4k_\mathrm{B}N}
\label{eq:Temp_pred}
\end{equation}
$N$ and $k_\mathrm{B}$ are the number of active (mobile) atoms and the Boltzmann constant, respectively.
In either case the equilibrated ideal Al-AlN(0001) surfaces are then reloaded for each ion bombardment. A consecutive bombardment inherently leads to a drifting surface state. A quantification and analysis of the overlaying effects of different ion energies, ion species, and surface states is out of scope for this work, however. 

To simulate the ion bombardment, the time step is eventually lowered starting at 0.25 fs to secure a maximum change of the atoms' kinetic energy and displacement below 0.1 eV and 0.01 \r A, respectively. The simulation time $t$ per impingement is defined by $t=2~\mathrm{ps}+3({\sqrt{E_\mathrm{ion}}-\sqrt{10~\mathrm{eV}}})/({\sqrt{100~\mathrm{eV}}-\sqrt{10~\mathrm{eV}}})~\mathrm{ps}$, to reflect the longer relaxation time for higher ion energies. 2 ps ($E_\mathrm{ion}$=10 eV) to 5 ps ($E_\mathrm{ion}$=100 eV) long RMD simulations are performed for each energy.

\begin{figure*}
\centering
\includegraphics[width=16cm]{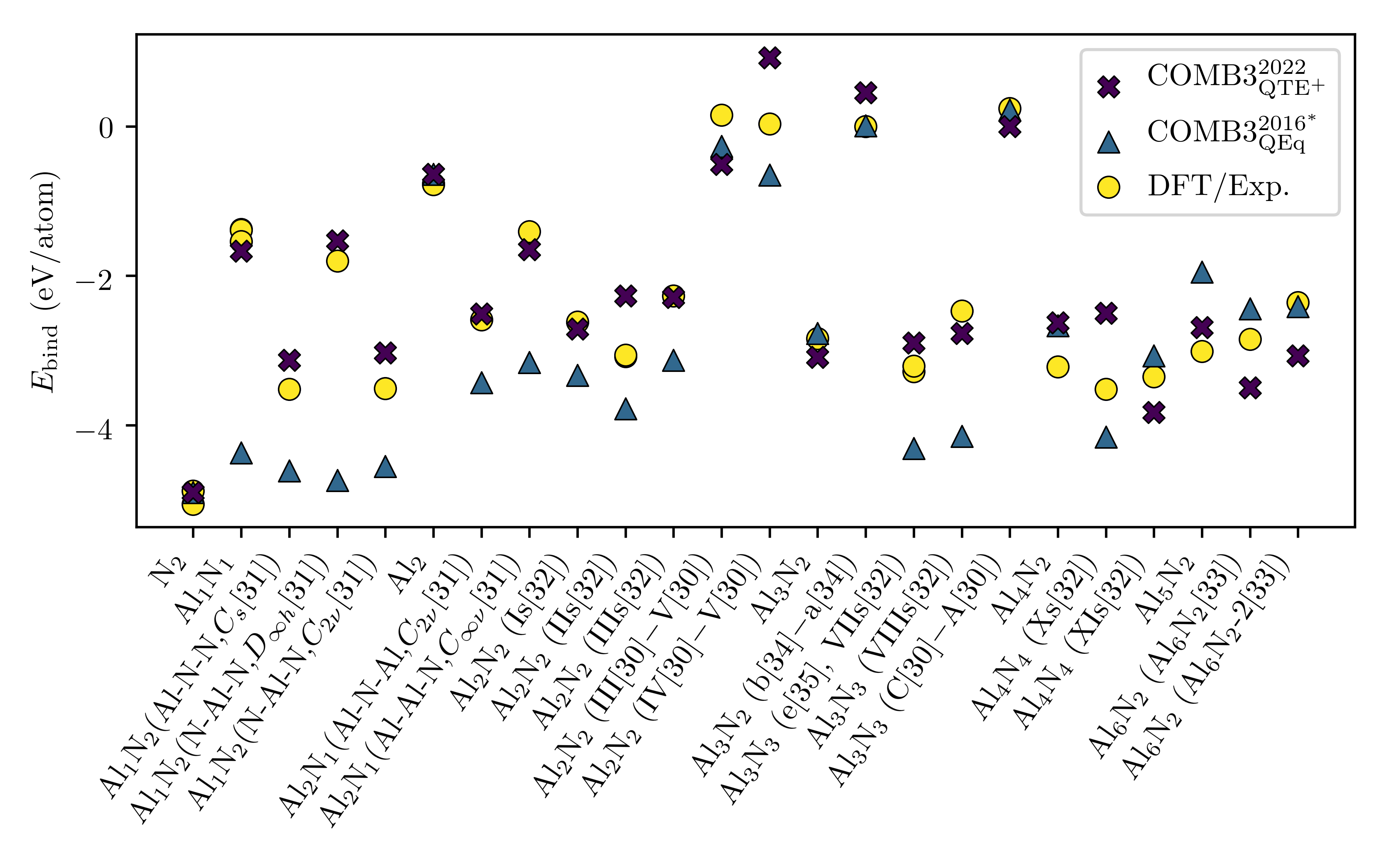}
\caption{Binding energies $E_\mathrm{bind}$ (eV/atom) of Al$_x$N$_y$ cluster ($x\leq6$, $y\leq4$) computed with COMB3$_\mathrm{QTE^+}^{2022}$ and COMB3$_\mathrm{QEq}^{2016^*}$ are compared with DFT and experimental findings \cite{herzberg_molecular_1950,nayak_evolution_1998,costales_density_2003,kandalam_first_2000,belbruno_structure_1999,costales_chemical_2002,song_stable_2002,kandalam_theoretical_2001}. The references are broken down in the Supplementary Table~S1. Labels in brackets allow for a structure identification in the provided reference when necessary. The subtraction of two labels indicates a relative binding energy.}
\label{fig:cluster} 
\end{figure*}

\section*{Results}

First, the results of the COMB3$_\mathrm{QEq}^{2016^*}$ and the revised COMB3$_\mathrm{QTE^+}^{2022}$ AlN potential on the reference set outlined previously are compared to density functional theory (DFT) and experimental (exp.) findings. Second, an examplary study of the ion bombardment induced flux emission and the damage formation is presented. 

\subsection*{Performance on reference set}

The fidelity of modeling Al$_x$N$_y$ cluster is assessed at first. The (relative) binding energies for clusters with up to eight atoms ($x\leq6$, $y\leq4$) are shown in Fig.~\ref{fig:cluster}. Dissociation energies are converted to binding energies by imposing a binding energy of -5.06 eV/atom and -0.77 eV/atom for N$_2$ and Al$_2$, respectively \cite{nayak_evolution_1998}. COMB3$_\mathrm{QEq}^{2016^*}$ describes the differentiation between exothermic and endothermic binding energies correctly. However, the absolute values are found to deviate significantly from the reference values. In particular the binding energy of Al$_1$N$_1$ are approximately three times too high. This could introduce exaggerated sputtering of such monomers due to ion bombardment. An approximately threefold overestimation is also found for Al$_1$N$_2$ (N-Al-N,$D_{\infty h}$ \cite{kandalam_first_2000}). In general it is found that COMB3$_\mathrm{QEq}^{2016^*}$ describes most clusters with up to three Al and N atoms to be too energetically favorable. In constrast, COMB3$_\mathrm{QTE^+}^{2022}$ models almost all clusters appropriately. The systematic overestimation is overcome. A larger deviation of $\geq$ 1 eV is only found for Al$_4$N$_4$ (Xs \cite{belbruno_structure_1999}).

\begin{table}[ht]
\centering
\begin{tabular}{ |l|c|c|c|c|}
\hline
Wurtzite & COMB3$_\mathrm{QTE^+}^{2022}$ & COMB3$_\mathrm{QEq}^{2016^*}$& COMB3$_\mathrm{QEq}^{2016}$  & DFT/Exp. \\
\hline
$a$ (\r A) 												& 3.14 & 3.10 & 3.11\cite{choudhary_dynamical_2016} & 3.11\cite{yim_epitaxially_1973,tanaka_morphology_1997,zoroddu_first-principles_2001}, 3.13\cite{holec_surface_2012,zhang_mechanism_2007} \\
\hline
$c/a$													& 1.60 & 1.62 & 1.62\cite{choudhary_dynamical_2016} & 1.60\cite{yim_epitaxially_1973,zhang_mechanism_2007,tanaka_morphology_1997,holec_surface_2012,zoroddu_first-principles_2001} \\
\hline
$C_{11}$ (GPa)											& 510 & 453 & 463\cite{choudhary_dynamical_2016} & 345\cite{wright_elastic_1997}, 464\cite{ruiz_electronic_1994} \\
\hline
$C_{12}$ (GPa)											& 95 & 87 & 92\cite{choudhary_dynamical_2016} & 125\cite{wright_elastic_1997}, 149\cite{ruiz_electronic_1994} \\
\hline
$C_{13}$ (GPa)											& 115 & 103 & 104\cite{choudhary_dynamical_2016} & 116\cite{ruiz_electronic_1994}, 120\cite{wright_elastic_1997} \\
\hline
$C_{33}$ (GPa)											& 494 & 463 & 437\cite{choudhary_dynamical_2016} & 395\cite{wright_elastic_1997}, 409\cite{ruiz_electronic_1994} \\
\hline
$C_{44}$ (GPa)											& 223 & 191 & 194\cite{choudhary_dynamical_2016} & 118\cite{wright_elastic_1997}, 128\cite{ruiz_electronic_1994} \\
\hline
$\Delta H_\mathrm{f}$ (eV/atom)							& -1.49 & -1.30 & -1.50\cite{choudhary_dynamical_2016} & -1.49\cite{holec_surface_2012}, -1.50\cite{zhang_mechanism_2007}, -1.57\cite{zoroddu_first-principles_2001} \\
\hline
\end{tabular}
\caption{\label{table: Wurzite}Wurtzite AlN properties obtained with COMB3$_\mathrm{QTE^+}^{2022}$ and COMB3$_\mathrm{QEq}^{2016^*}$ are compared to the COMB3$_\mathrm{QEq}^{2016}$ AlN publication, and  DFT as well as experimental findings.}
\end{table}

The considered wutzite AlN properties are listed in Table~\ref{table: Wurzite}. The results of COMB3$_\mathrm{QTE^+}^{2022}$ and COMB3$_\mathrm{QEq}^{2016^*}$ are compared with DFT and experimental findings. Both potentials describe the lattice constants appropriately, but tend to overestimate the elastic constants $C_{11}$, $C_{33}$ and $C_{44}$. This bias is more pronounced for the AlN potential revised in this work. The heat of formation is modeled with high accuracy with COMB3$_\mathrm{QTE^+}^{2022}$, whereas the COMB3$_\mathrm{QEq}^{2016^*}$ potential underestimates this parameter by approximately 0.2 eV. 

\begin{figure}[ht]
\centering
\includegraphics[width=\linewidth]{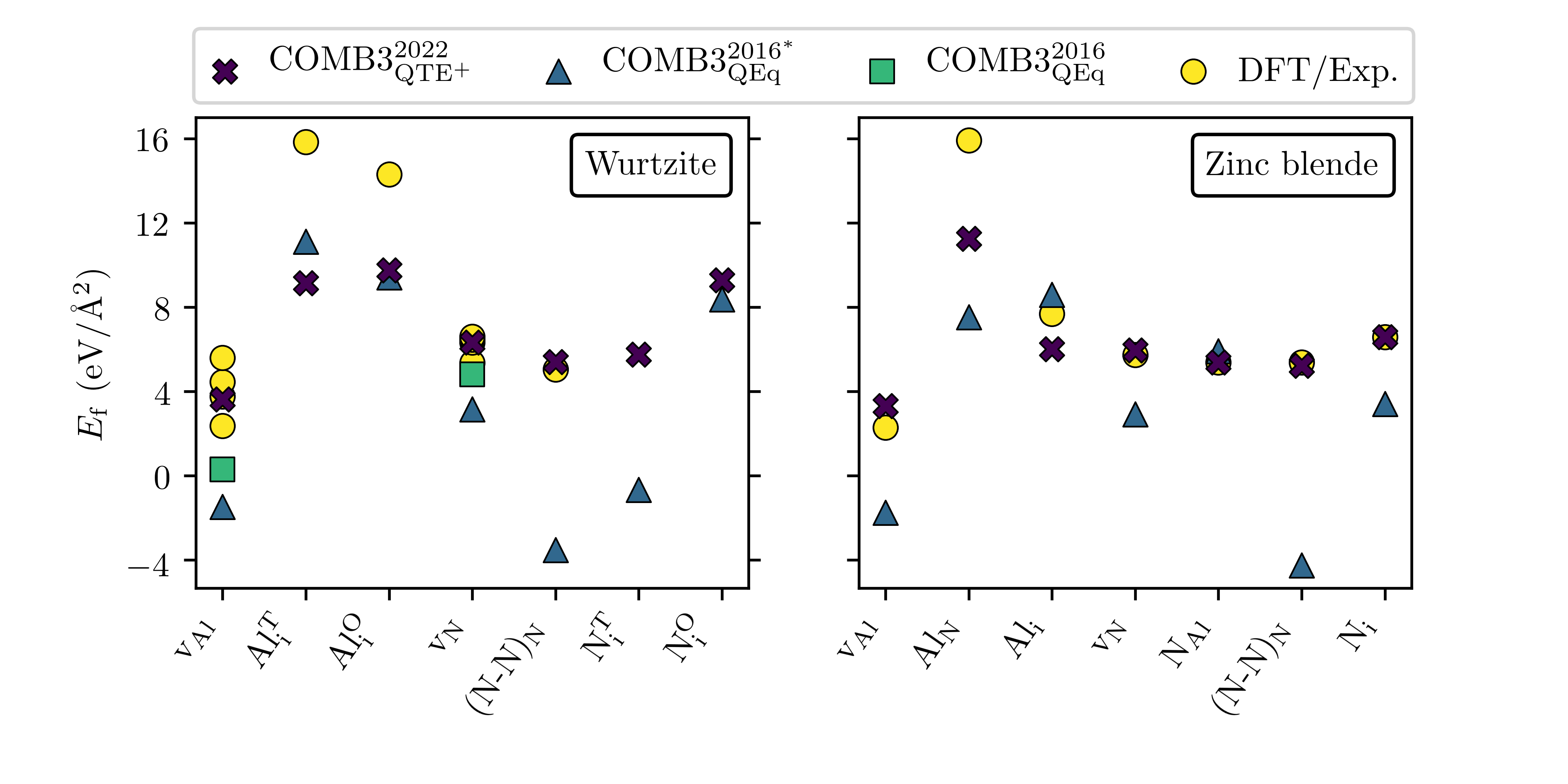}
\caption{AlN defect formation energies obtained with COMB3$_\mathrm{QTE^+}^{2022}$ and COMB3$_\mathrm{QEq}^{2016^*}$ are compared to the COMB3$_\mathrm{QEq}^{2016}$ AlN publication, and  DFT as well as experimental findings\cite{choudhary_dynamical_2016,hung_ab_2004,stampfl_theoretical_2002,gorczyca_theory_1999,szallas_characterization_2014}. The references are broken down in the Supplementary Table~S2 and Supplementary Table~S3.}
\label{fig:EDefects} 
\end{figure}

Furthermore, it struggles with the v$_\mathrm{Al}$ and the (N-N)$_\mathrm{N}$ point defect structures presented in Fig.~\ref{fig:EDefects}. The negative values are likely to cause an exaggerated number of vacant Al sites and incorporated N split interstitials during an ongoing ion bombardment. In contrast, COMB3$_\mathrm{QTE^+}^{2022}$ is found to describe this and the other point defects reasonably well. 

\begin{figure}[ht]
\centering
\includegraphics[width=\linewidth]{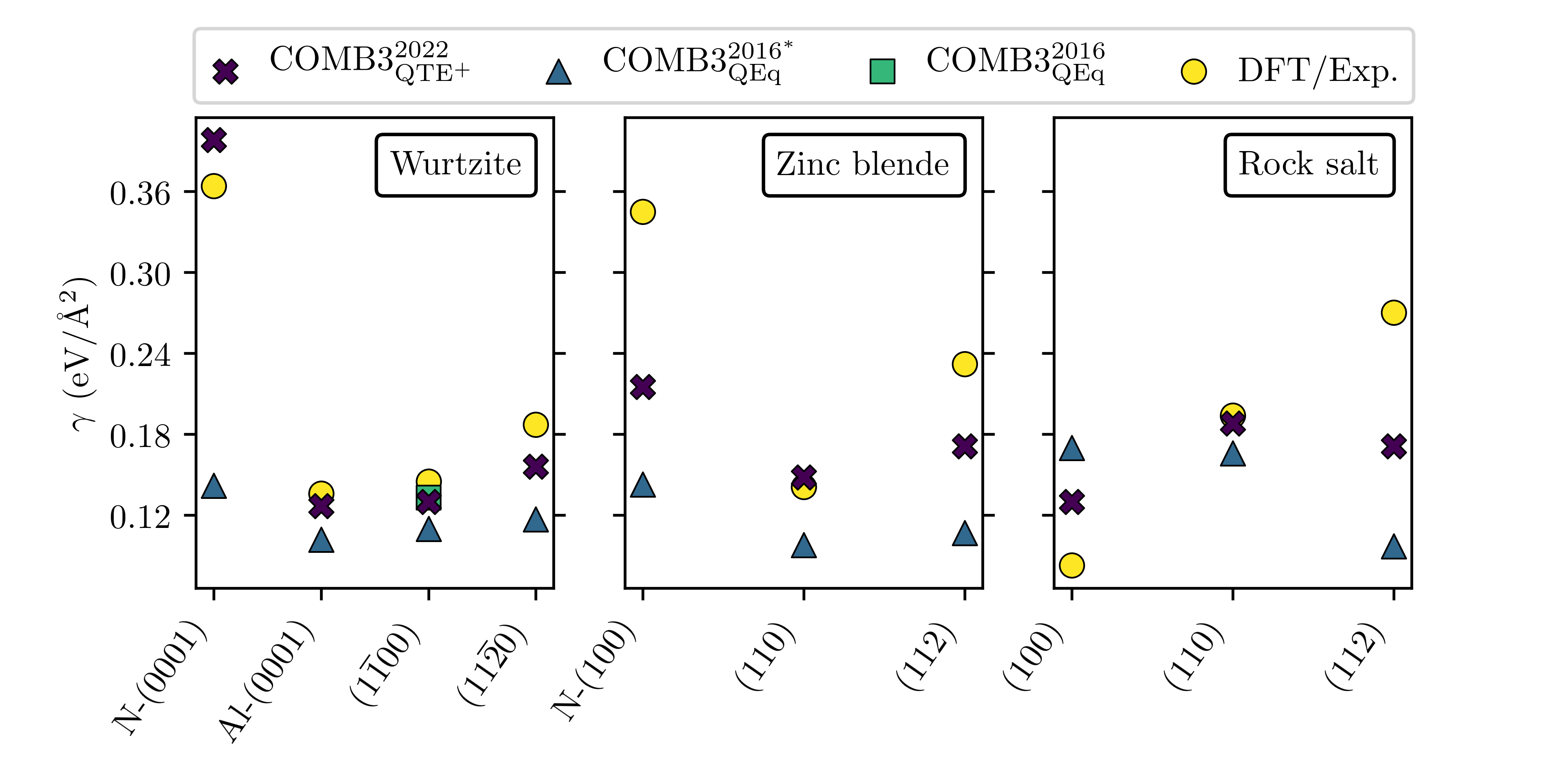}
\caption{AlN surface energies obtained with COMB3$_\mathrm{QTE^+}^{2022}$ and COMB3$_\mathrm{QEq}^{2016^*}$ are compared to the COMB3$_\mathrm{QEq}^{2016}$ AlN publication, and  DFT as well as experimental findings \cite{jindal_density_2009,choudhary_dynamical_2016,holec_surface_2012}. The references are broken down in the Supplementary Table~S2, Supplementary Table~S3, and Supplementary Table~S4.}
\label{fig:surfaceEs} 
\end{figure}

Both potentials describe the order of surface energies shown in Fig.~\ref{fig:surfaceEs} correctly (i.e., $\gamma_\mathrm{\text{Al-(0001)}} < \gamma_\mathrm{(1\overline{1}00)} < \gamma_\mathrm{(11\overline{2}0)} < \gamma_\mathrm{\text{N-(0001)}}$). Though the AlN potential revised in this work predict smaller deviations than its predecessor.

\begin{table}[ht]
\centering
\begin{tabular}{ |l|c|c|c|c|}
\hline
Zinc blende & COMB3$_\mathrm{QTE^+}^{2022}$ & COMB3$_\mathrm{QEq}^{2016^*}$& COMB3$_\mathrm{QEq}^{2016}$  & DFT/Exp. \\
\hline
$a$ (\r A) 												& 4.40 & 4.37 & & 4.37\cite{neumann_j_1995}, 4.40\cite{holec_surface_2012,zhang_mechanism_2007} \\
\hline
$C_{11}$ (GPa)											& 521 & 468 & & 304\cite{kim_elastic_1996}, 313\cite{shimada_first-principles_1998} \\
\hline
$C_{12}$ (GPa)											& 115 & 76 & & 152\cite{kim_elastic_1996}, 160\cite{shimada_first-principles_1998} \\
\hline
$C_{44}$ (GPa)											& 190 & 169 & & 192\cite{shimada_first-principles_1998}, 199\cite{kim_elastic_1996} \\
\hline
$\Delta H_\mathrm{f}$ (eV/atom)							& -1.33 & -1.17 & -1.38\cite{choudhary_dynamical_2016} & -1.46\cite{holec_surface_2012}, -1.47\cite{zhang_mechanism_2007}, -1.49\cite{zoroddu_first-principles_2001} \\
\hline
$\Delta H_\mathrm{rxn}^\mathrm{ZnB-WZ}$ (eV)							& 0.16 & 0.13 & 0.25\cite{choudhary_dynamical_2016} & 0.04\cite{holec_surface_2012}, 0.05\cite{choudhary_dynamical_2016} \\
\hline
\end{tabular}
\caption{\label{table: zinc blende}Zinc blende AlN properties obtained with COMB3$_\mathrm{QTE^+}^{2022}$ and COMB3$_\mathrm{QEq}^{2016^*}$ are compared to the COMB3$_\mathrm{QEq}^{2016}$ AlN publication, and  DFT as well as experimental findings.}
\end{table}

The considered properties for the zinc blende AlN structure are listed in Table~\ref{table: zinc blende}. The results of COMB3$_\mathrm{QTE^+}^{2022}$ and COMB3$_\mathrm{QEq}^{2016^*}$ are again compared to DFT and experimental findings. Both models describe the lattice constant and elastic constant $C_{44}$ accurately. Though both models also under- and overestimate $C_{12}$ and $C_{11}$, respectively. The reference heat of formation $\Delta H_\mathrm{f}$ and heat of reaction $\Delta H_\mathrm{rxn}$ match reasonably well.

Concerning the point defect structure shown in Fig.~\ref{fig:EDefects}, COMB3$_\mathrm{QEq}^{2016^*}$ describes only Al$_\mathrm{i}$ and N$_\mathrm{Al}$ accurately. For all other point defects, deviations of more than 2 eV are observed. The binding energy of vacant Al sites v$_\mathrm{Al}$ and N split interstitials (N-N)$_\mathrm{N}$ even deviate by more than 178 \% and 177 \%, respectively. This issue with these two point defect structures is also observed for the wurtzite structure, discussed in the preceding paragraph (see Table~\ref{table: Wurzite}). This repetition signifies a systematic problem of the COMB3$_\mathrm{QEq}^{2016^*}$ potential with v$_\mathrm{Al}$ and (N-N)$_\mathrm{N}$. The last is assumed to be caused by its nitrogen bond configuration resembling molecular nitrogen. In contrast, COMB3$_\mathrm{QTE^+}^{2022}$ describes all point defect structures with high fidelity.

Both models describe the order of surface energies shown in Fig.~\ref{fig:surfaceEs} correctly (i.e., $\gamma_\mathrm{(110)} < \gamma_\mathrm{(112)} < \gamma_\mathrm{\text{N-(100)}}$). Though the revised COMB3 AlN potential bests the previous one at closing the gap to the reference values. 

\begin{table}[ht]
\centering
\begin{tabular}{|l|c|c|c|c|}
\hline
Rock salt & COMB3$_\mathrm{QTE^+}^{2022}$ & COMB3$_\mathrm{QEq}^{2016^*}$ & COMB3$_\mathrm{QEq}^{2016}$ & DFT/Exp. \\
\hline
$a$ (\r A) 												& 4.06 & 4.15 & &  4.06\cite{siegel_ab_2006}, 4.07\cite{holec_surface_2012,zhang_mechanism_2007} \\
\hline
$\Delta H_\mathrm{f}$ (eV/atom)							& -1.32 & -0.17 &  -0.40\cite{choudhary_dynamical_2016} & -1.30\cite{holec_surface_2012}, -1.31\cite{zhang_mechanism_2007} \\
\hline
$\Delta H_\mathrm{rxn}^\mathrm{NaCl-WZ}$ (eV)							& 0.35 & 2.27 & 2.2\cite{choudhary_dynamical_2016} & 0.36\cite{holec_surface_2012}, 0.4\cite{choudhary_dynamical_2016} \\
\hline
\end{tabular}
\caption{\label{table: rock salt}Rock salt AlN properties obtained with COMB3$_\mathrm{QTE^+}^{2022}$ and COMB3$_\mathrm{QEq}^{2016^*}$ are compared to the COMB3$_\mathrm{QEq}^{2016}$ AlN publication, DFT, and experimental references.}
\end{table}

The properties considered for the rock salt structure are listed in Table~\ref{table: rock salt}. COMB3$_\mathrm{QEq}^{2016^*}$ overestimates the lattice constants and fails to resolve the heat of formation and the heat of reaction. In contrast, COMB3$_\mathrm{QTE^+}^{2022}$ describes the lattice constant, the heat of formation, and the heat of reaction accurately. 

However, both models are not capable of describing the correct order of surface energies depicted in Fig.~\ref{fig:surfaceEs}, whereas COMB3$_\mathrm{QTE^+}^{2022}$ provides a smaller mean absolute error (i.e., 0.051 eV/\r A$^2$) than COMB3$_\mathrm{QEq}^\mathrm{2016^*}$ (i.e., 0.096 eV/\r A$^2$).

The heat of reaction of CsCl relative to the wurtzite structure $\Delta H_\mathrm{rxn}^\mathrm{CsCl-WZ}$ for COMB3$_\mathrm{QTE^+}^{2022}$, COMB3$_\mathrm{QEq}^{2016^*}$,  COMB3$_\mathrm{QEq}^{2016}$, and DFT reference are 4.02 eV, 1.37 eV, 1.37 eV, and 4.03 eV, respectively \cite{choudhary_dynamical_2016}. Hence, COMB3$_\mathrm{QTE^+}^{2022}$ provides a more accurate description of the phase stability. Relaxed atom structures resemble the referenced configurations for all herein considered cases by means of bond lengths and bond angles (not shown).

\subsection*{Exemplary study: Ion impingement on AlN(0001)}

\begin{figure}[ht]
\centering
\includegraphics[width=\linewidth]{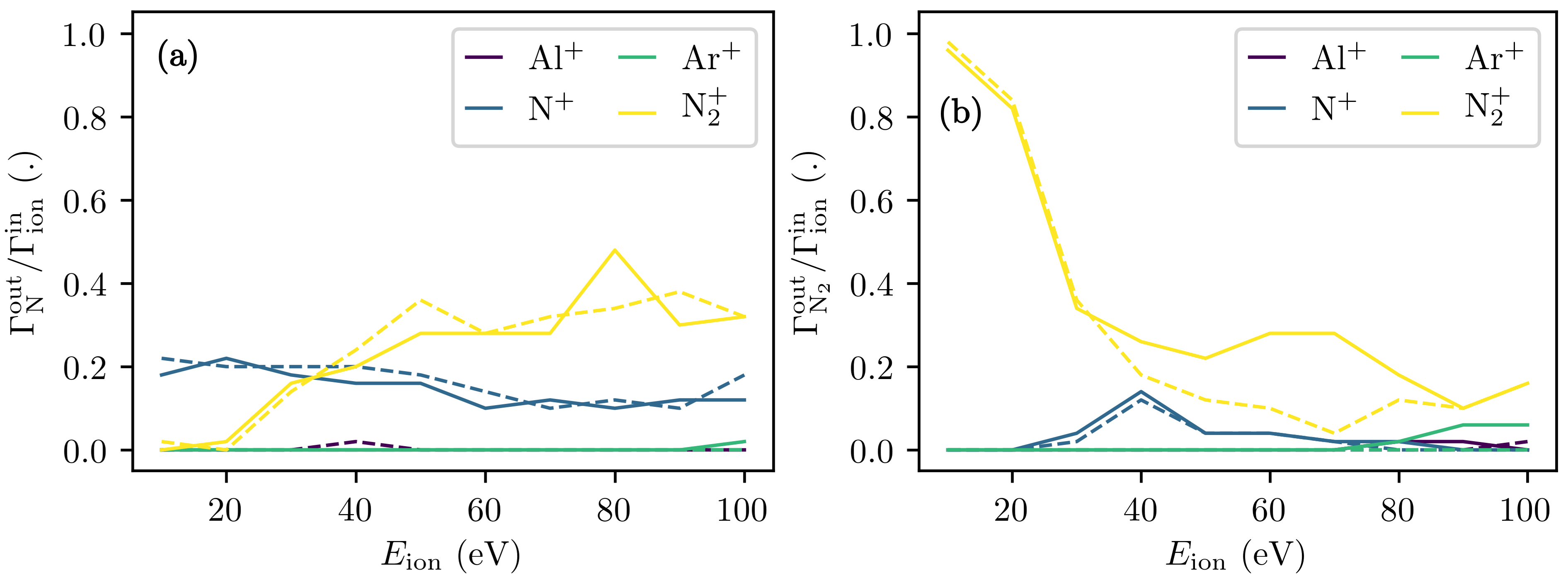}
\caption{\textbf{(a)} Nitrogen flux $\Gamma_\mathrm{N}^\mathrm{out}$ and \textbf{(b)} molecular nitrogen flux $\Gamma_{\mathrm{N}_2}^\mathrm{out}$ per incident ion flux $\Gamma_\mathrm{ion}^\mathrm{in}$ are presented as a function of the ion energy $E$ and ion species (i.e., Al$^+$, N$^+$, Ar$^+$, N$_2^+$). Solid and dashed lines indicate surface initially equilibrated at 300 K and relaxed at 0 K, respectively.}
\label{fig:NF} 
\end{figure}

\begin{figure}[ht]
\centering
\includegraphics[width=8 cm]{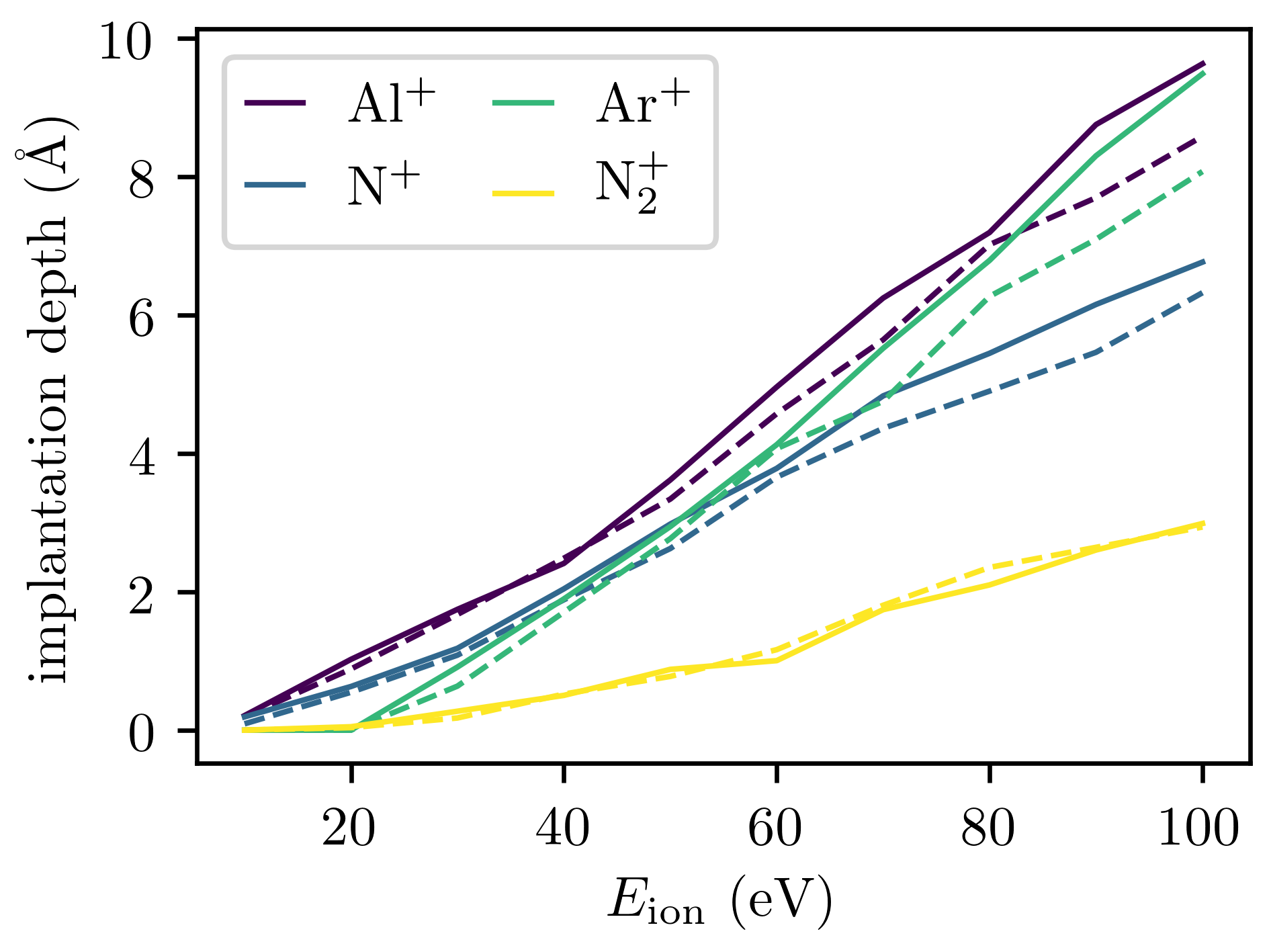}
\caption{Implantation depth is presented as a function of the ion energy $E$ and ion species (i.e., Al$^+$, N$^+$, Ar$^+$, N$_2^+$). Solid and dashed lines indicate surfaces initially equilibrated at 300 K and relaxed at 0 K, respectively.}\label{fig:impl_depth} 
\end{figure}

In the following, the irradiation of an ideal wurtzite AlN(0001) surface by Al$^+$, N$^+$, Ar$^+$, and N$_2^+$ ions is investigated. The emitted nitrogen flux $\Gamma_\mathrm{N}^\mathrm{out}$ per incident ion flux $\Gamma_\mathrm{ion}^\mathrm{in}$ is displayed in Fig.~\ref{fig:NF} \textbf{(a)}. As evident, a reflection probability for N$^+$ of 20.0 \% for $E_\mathrm{ion}=10$ eV decreases down to 12.0 \% for $E_\mathrm{ion}\geq60$ eV (root-mean-square deviation: 2.2 \%). Above 20 eV a conversion of incident nitrogen ions and surface atoms to molecular nitrogen is observed. This process peaks at 40 eV as depicted in Fig.~\ref{fig:NF} \textbf{(b)}. An incident nitrogen ion eventually hits a surface nitrogen atom and forms a nitrogen split interstitial (N-N)$_\mathrm{N}$ at the surface for a brief moment until the inverted momentum (now in surface normal direction) leads to the emission of the just formed (N-N)$_\mathrm{N}$ as a N$_2$ molecule. For even higher ion energies (i.e., $E_\mathrm{ion}>40$ eV) this process becomes less relevant due the more likely and deeper ion implantation as depicted in Fig.~\ref{fig:impl_depth}.

\begin{figure}[ht]
\centering
\includegraphics[width=\linewidth]{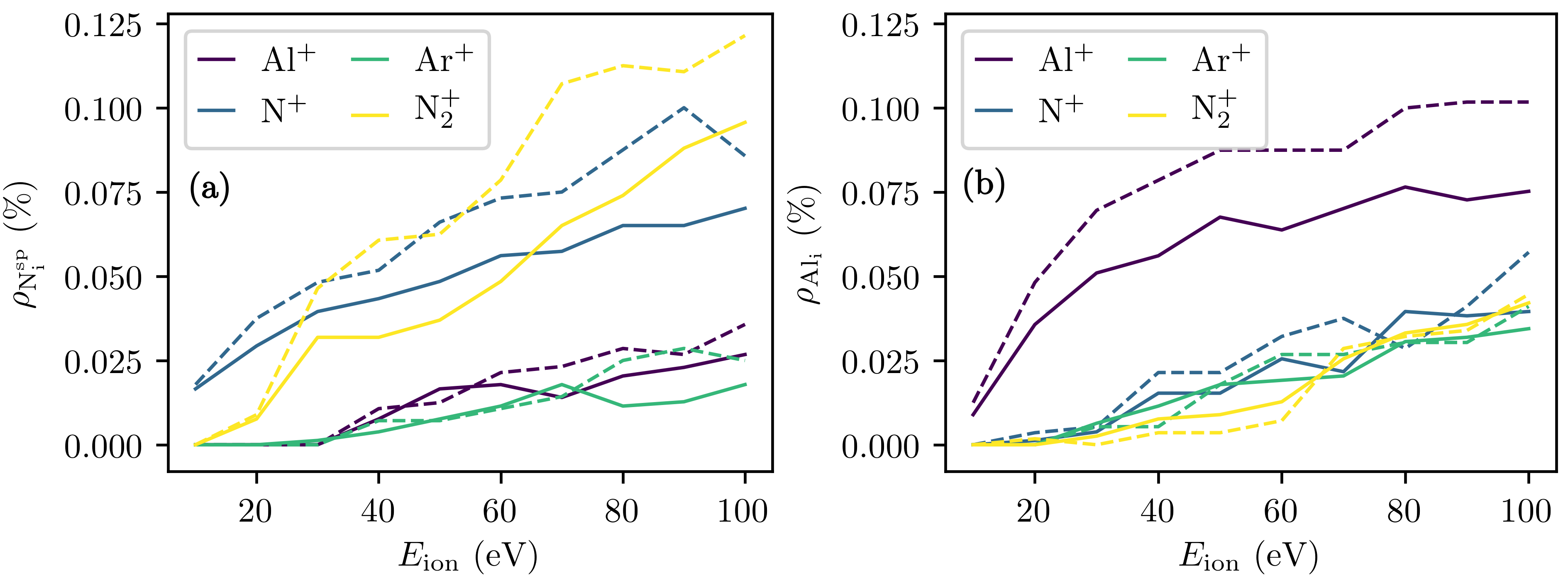}
\caption{\textbf{(a)} Nitrogen split interstitial population $\rho_\mathrm{N_i^\mathrm{sp}}$ and \textbf{(b)} Aluminum interstitial population $\rho_\mathrm{Al_i}$ are presented as a function of the ion energy $E$ and ion species (i.e., Al$^+$, N$^+$, Ar$^+$, N$_2^+$). Solid and dashed lines indicate surfaces initially equilibrated at 300 K and relaxed at 0 K, respectively.}
\label{fig:I} 
\end{figure}

The same process also leads to an increase in the emission of molecular nitrogen for incident N$_2^+$ that is split up into two nitrogen atoms at the surface for ion energies of 40-80 eV. This process is peaked at 60-70 eV. The broadening and shift to higher ion energies is reasoned by a decreased implantation depth (see Fig.~\ref{fig:impl_depth}). This is due the initially shared and eventually separated momentum. The implantation depths of N$^+$ and N$_2^+$ for approximately 40 eV and 60-70 eV, respectively, are indeed comparable. Hence, the impingement of N$_2^+$ intrinsically affects a region closer to the surface which facilitates the formation and eventual emission of split interstitials as molecular nitrogen. 

The incoming flux of N$_2^+$ contributes most significantly to the outgoing nitrogen atom flux for $E_\mathrm{ion}>20$ eV due to a splitting and subsequent reflection of one N atom at the surface. This process increases only slightly for energies $E_\mathrm{ion}>40$ eV. The remaining second N atom may be implanted, stick, reflected, or leave the surface with another N surface atom as an N$_2$ molecule. The last process contributes most substantially to N$_2$ emission for energies of 30-80 eV (sticking is dominant at lower energies and implantation is dominant at higher energies).

\begin{figure}[ht]
\centering
\includegraphics[width=8 cm]{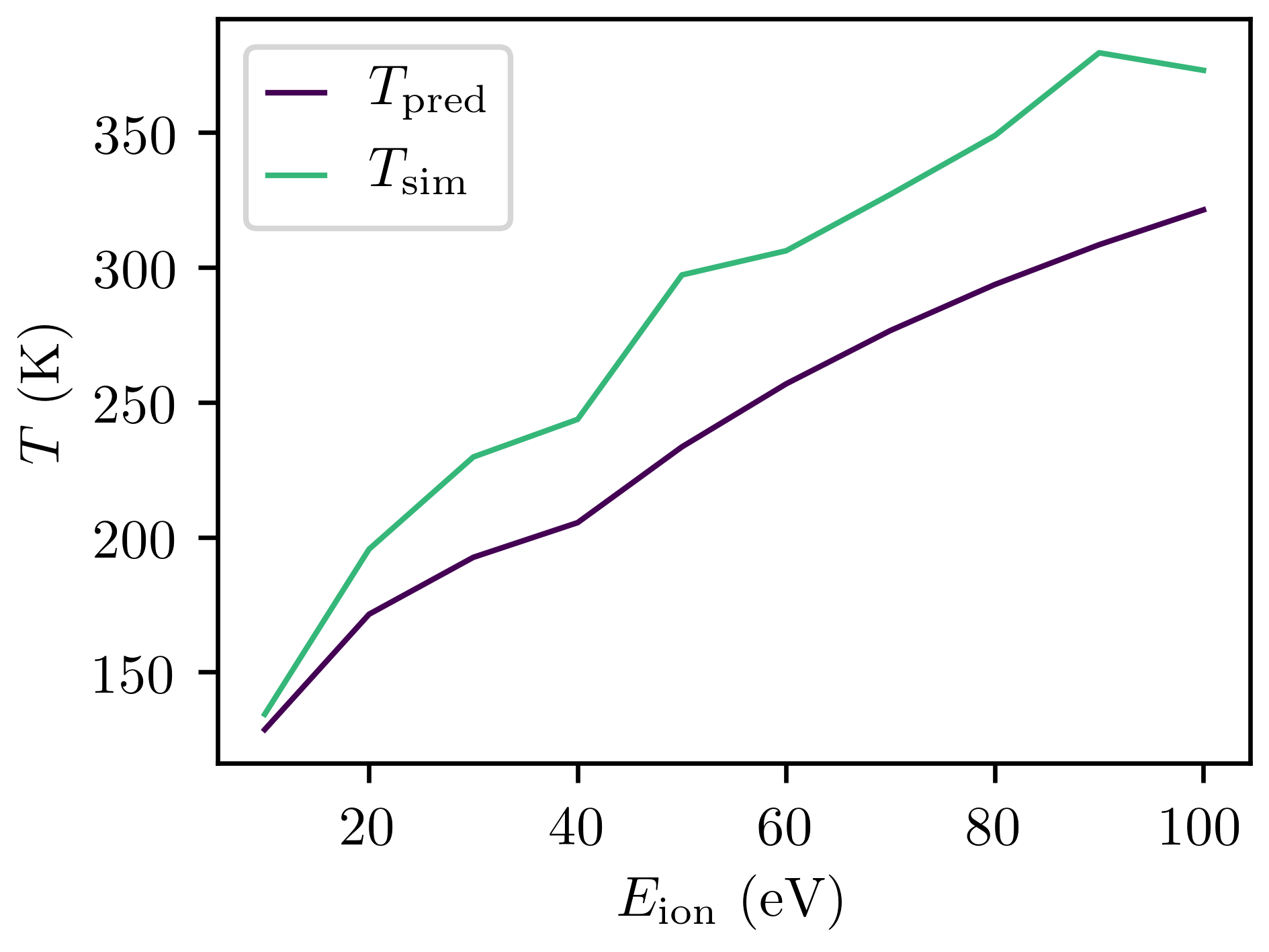}
\caption{Theoretically predicted (see equation \eqref{eq:Temp_pred}) and by simulation obtained surface slab temperatures $T_\mathrm{pred}$ and $T_\mathrm{sim}$ are presented as function of the ion energy $E$.}
\label{fig:Temp} 
\end{figure}

Only a negligible number of Al ions and Al surface atoms are reflected and sputtered, respectively (not shown). This is due to high sticking probabilities and small sputtering yields in the considered ion energy range (i.e., 10-100 eV). The incorporated Al interstitials Al$_\text{i}$ are quantified by means of their defect population as shown in Fig.~\ref{fig:I} \textbf{(b)}. Up to approximately 30 eV, solely the implantation of impinging Al$^+$ ions contributes to the Al interstitial population. For higher ion energies, forward sputtering (peening) of Al surface atoms leads to a steadily increasing number of Al interstitials for any in this work considered ion species. 

All Ar ions are reflected for ion energies equal to 10-20 eV. At 30 eV 76 \% of the incident Ar ions are still reflected. The reflection probability decreases linearly with increasing ion energies down to 29 \% at $E_\mathrm{ion}=100$ eV. The N and Al vacancy populations are increased for higher ion energies due to the forward sputtering (peening) of N and Al atoms, respectively (see Supplementary Fig.~S6). Only a negligible amount of other point defect populations (i.e., $\rho_\mathrm{Al_N}$, $\rho_\mathrm{N_{Al}}$, $\rho_\mathrm{Ar_{Al}}$, $\rho_\mathrm{Ar_N}$) are observed (not shown).

Starting off with a relaxed surface at 0 K (see equation \eqref{eq:Temp_pred}) provides comparable results for nearly all properties (e.g., Ar reflection probability, implantation depth). However, a noticeable deviation is observed for the emitted molecular nitrogen flux $\Gamma_{\mathrm{N}_2}^\mathrm{out}$ as a function of the incident N$_2^+$ ion flux $\Gamma_\mathrm{ion}^\mathrm{in}$ for $40$ eV$<E_\mathrm{ion}<80$ eV, as displayed in Fig.~\ref{fig:NF} \textbf{(b)}. The described process of splitting, forming split interstitials, and leaving the system as N$_2$ molecules is weakened when starting with a relaxed surface at 0 K. A too large proportion of the ions' kinetic energy is transferred to the surface atoms which eventually hinders this process for this ion energy range. For higher ion energies (i.e., 90-100 eV) this issue is resolved due to the deeper implantation depth (see Fig.~\ref{fig:NF} \textbf{(b)} and Fig.~\ref{fig:impl_depth}). Moreover, it is found that starting with a relaxed surface at 0 K promotes the ion implantation and generation of corresponding interstitials (see Al$^+$ in Fig.~\ref{fig:I} \textbf{(b)}, N$^+$ in Fig.~\ref{fig:I} \textbf{(a)}, and N$_2^+$ in Fig.~\ref{fig:I} \textbf{(a)}). It has a minor effect on the ion bombardment induced Frenkel pair generation. 

The corresponding surface slab temperature $T$ is depicted in Fig.~\ref{fig:Temp}. $T_\mathrm{sim}$ and $T_\mathrm{pred}$ are the simulated mean temperature and its lower bound as predicted by theory (see equation \eqref{eq:Temp_pred}). A good agreement between theory and simulation is observed. The first may be used to design studies starting with a relaxed surface at 0 K and, hence, bypassing the burden of equilibrating a surface at a specific temperature. However, the provided theoretical estimate should be used with care when the initial atom configuration is high in potential energy (relative to its ground state). 

\section*{Conclusions}

GARFfield with its new extensions allowed for the revision of the COMB3$_\mathrm{QEq}^{2016}$ AlN potential. COMB3$_\mathrm{QTE^+}^{2022}$ is proposed to enable RMD studies of the wurtzite and zinc blende phase AlN plasma-surface interactions. The potential overcomes manifold shortcomings for the intended application (e.g., negative defect formation energies for v$_\mathrm{Al}$ and (N-N)$_\mathrm{N}$) of the COMB3$_\mathrm{QEq}^{2016}$ potential. The excellent agreement of the results obtained using the COMB3$_\mathrm{QTE^+}^{2022}$ potential with experimental references and DFT based computations indicates that the damage generation (e.g., Frenkel pair) during an ongoing ion bombardment is appropriately described. High-energy collisions are taken into account by merging (tapering) the COMB3$_\mathrm{QTE^+}^{2022}$ with the ZBL potential. The decent descriptions of most Al$_x$N$_y$ clusters with up to eight atoms (i.e., $x\leq6$, $y\leq4$) demonstrate that the COMB3$_\mathrm{QTE^+}^{2022}$ potential may be considered for systems involving under-coordinated atoms (e.g., atop surfaces). Moreover, it is expected to describe more reliably the relative amounts of different sputtered species (e.g., N, N$_2$, Al, Al$_1$N$_1$, Al$_2$).

The COMB3$_\mathrm{QTE^+}^{2022}$ may, however, be used with care when studying AlN in the rock salt phase. The potential is found to describe the lattice parameter and heat of formation (reaction) with high accuracy. But the correct order of surface energies could not be reproduced and, most importantly, only a small set reference properties is considered for this phase.

An example study on the effect of the isolated ion bombardment (i.e., Al$^+$, N$^+$, Ar$^+$, N$_2^+$) onto the ideal wurtzite Al-AlN(0001) surface was conducted. An analysis of the individual trajectories, particle emissions, and point defect formations for singular impingement events revealed that N$^+$ and N$_2^+$ may form N split interstitials with surface N atoms. These eventually leave as N$_2$ for ion energies in the range of 30-50 eV and 40-80 eV, respectively. The broadening and shift to higher energies for N$_2^+$ is explained by the initially shared and finally separated momentum. This process is weakened when the surfaces are initially not equilibrated at the 300 K, but relaxed at 0 K. In addition, in this case the incorporation of the incident ions is facilitated. Notably, most other properties (e.g., Frenkel pair populations, implantation depths) are found to be comparable. It is argued that such a procedure may be of interest for applications where its shortcomings are less critical, but the reduced computational costs are of importance (e.g., for parameters screenings). 

Self-adaptive evolution strategies are implemented into the GARFfield framework and allow for a more versatile interaction potential optimization. A continuous transition or alternation between genetic algorithm and evolution strategy can be readily applied, providing a more versatile evolutionary algorithm in GARFfield. Further extensions such as surface energy calculations or replacing the COMB with the COMB3 potential enhance the application possibilities.


\section*{Acknowledgement}

The authors thank Dr. Andres Jaramillo-Botero from California Institute of Technology, Division of Chemistry and Chemical Engineering for the permission to use the GARFfield simulation software. Funded by the Deutsche Forschungsgemeinschaft (DFG, German Research Foundation) – Project-ID 138690629 (TRR 87) and Project-ID 434434223 (SFB 1461).


\section*{Data Availability}

The revisited COMB3 AlN parameterization that can be utilized in combination with LAMMPS and data that support the findings of this study are available from the corresponding author upon reasonable request.


\section*{ORCID iDs}
\noindent
T. Gergs: \url{https://orcid.org/0000-0001-5041-2941}\\
T. Mussenbrock: \url{https://orcid.org/0000-0001-6445-4990} \\
J. Trieschmann: \url{https://orcid.org/0000-0001-9136-8019}


\section*{Author contributions statement}
T.G. prepared the data sets, implemented the methodology, conducted the simulations, analysed the results, and prepared the first draft of the manuscript. T.G. and J.T. developed the methodological concept. J.T. and T.M. acquired the funding, and administered the project. All authors contributed with interpreting and discussing the results as well as manuscript writing and revision.


\section*{Competing interests}
The authors declare no competing interests.


\section*{Additional information}
\textbf{Supplementary Information} The online version contains supplementary material.


\clearpage
\bibliography{./references.bib}


\end{document}


\section*{Supplementary Methods}

\begin{figure}[ht]
\centering
\includegraphics[width=\linewidth]{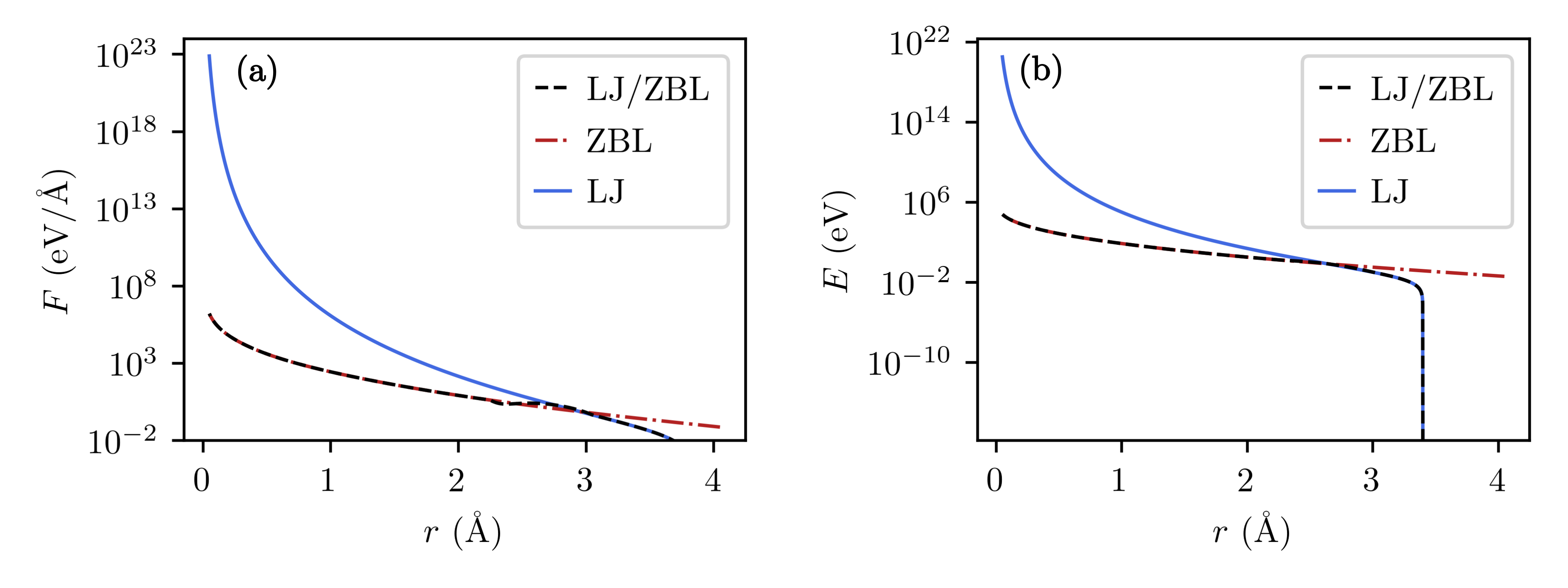}
\caption{\textbf{(a)} Force $F$ between and \textbf{(b)} potential energy $E$ of a pair of Ar atoms.  Both properties are presented as a function of the interatomic distance $r$.}
\label{fig:ArAr} 
\end{figure}

The implementation of the LJ/ZBL tapering is verified by comparing the forces $F$ and potential energies $E$ of the combined LJ/ZBL with the individual LJ as well as ZBL potential for a pair of Ar atoms. $F$ and $E$ are shown as a function of the interatomic distance in Fig.~\ref{fig:ArAr} \textbf{(a)} and \textbf{(b)}, respectively. 

\begin{figure}[ht]
\centering
\includegraphics[width=\linewidth]{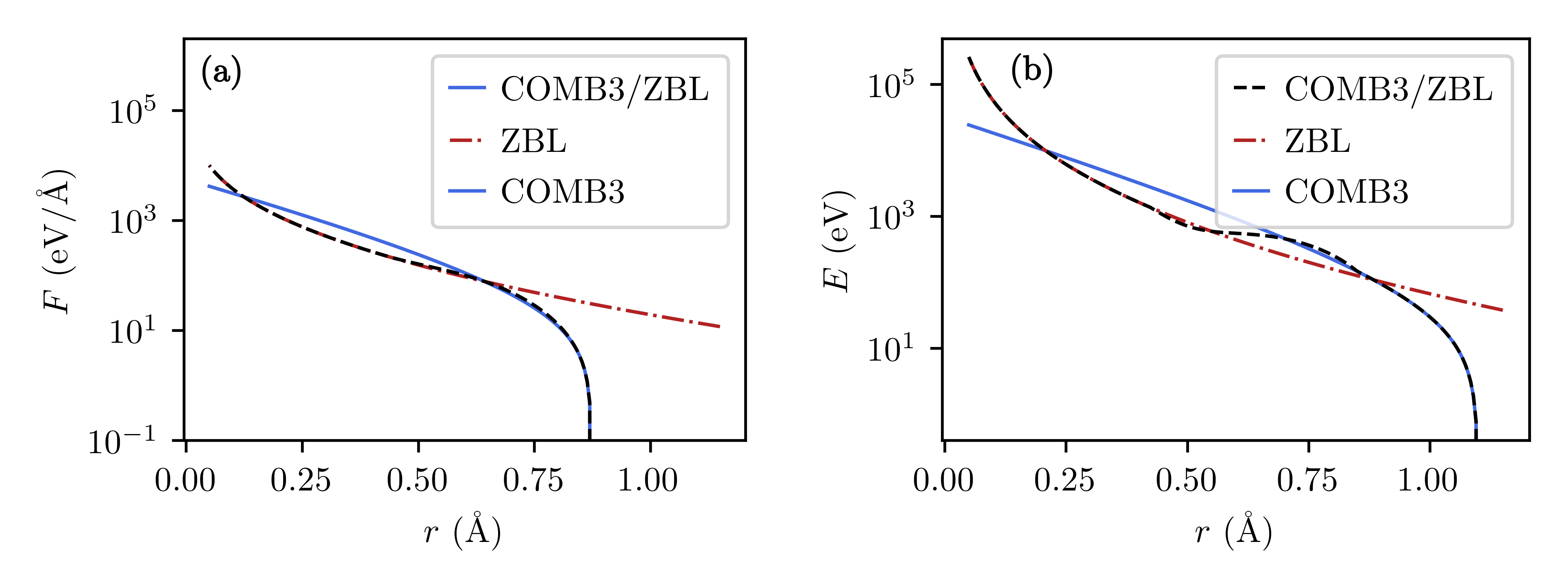}
\caption{\textbf{(a)} Force $F$ between and \textbf{(b)} potential energy $E$ of a pair of nitrogen atoms (molecular nitrogen). Both properties are presented as a function of the interatomic distance $r$.}
\label{fig:NNMol} 
\end{figure}

The implementation of the COMB3/ZBL tapering is verified by comparing the forces $F$ and potential energies $E$ of the combined COMB3/ZBL with the individual COMB3 as well as ZBL potential for a variety of cases. First, molecular nitrogen is considered. $F$ and $E$ are shown as a function of the interatomic distance in Fig.~\ref{fig:NNMol} \textbf{(a)} and \textbf{(b)}, respectively. 

\begin{figure}[ht]
\centering
\includegraphics[width=\linewidth]{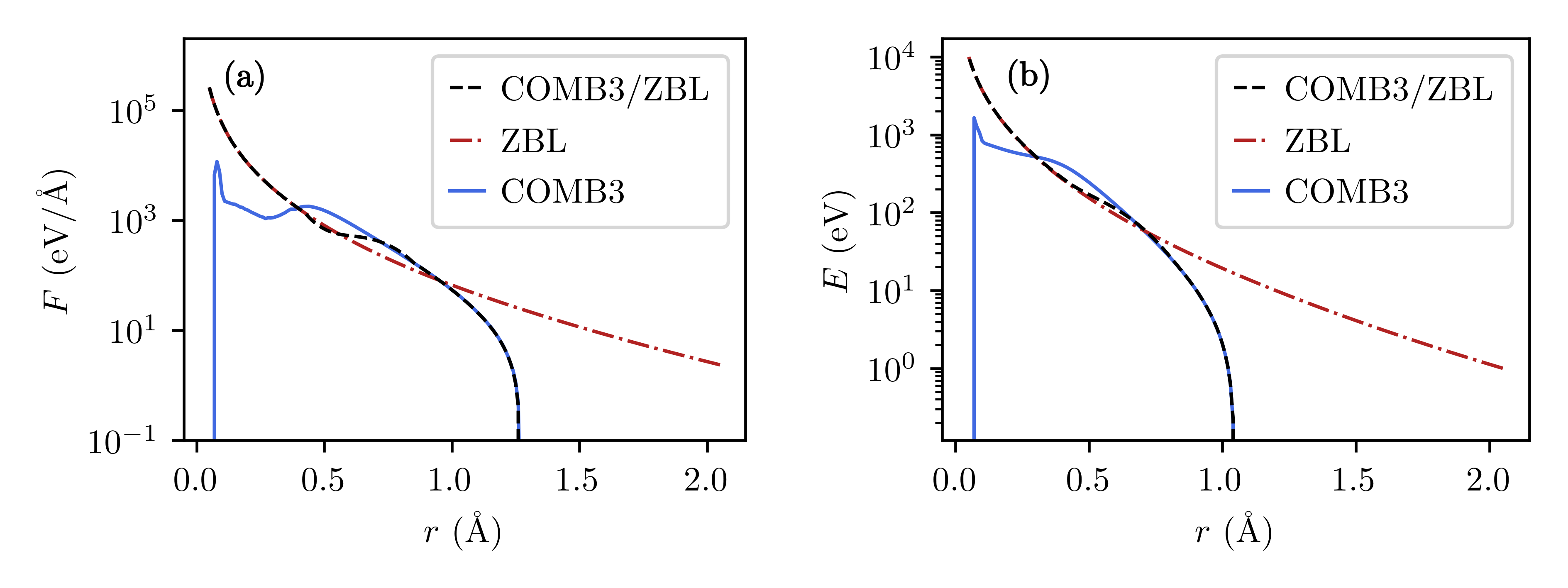}
\caption{\textbf{(a)} Force $F$ between and \textbf{(b)} potential energy of a N adatom atop a N wurtzite AlN(0001) surface atom. The contribution of the remaining surface is excluded. Both properties are presented as a function of the interatomic distance $r$.}
\label{fig:NN} 
\end{figure}

Second, a N atom on top the wurtzite AlN(0001) surface is lowered until it overlaps with a N surface atom. This simulation is performed twice. In the second run, interactions between the N adatom and N surface atom are excluded. The difference of both runs are used to obtain the N-N pair interaction for this many-body scenario. $F$ and $E$ are shown as a function of the interatomic distance in Fig.~\ref{fig:NN} \textbf{(a)} and \textbf{(b)}, respectively.

\begin{figure}[ht]
\centering
\includegraphics[width=\linewidth]{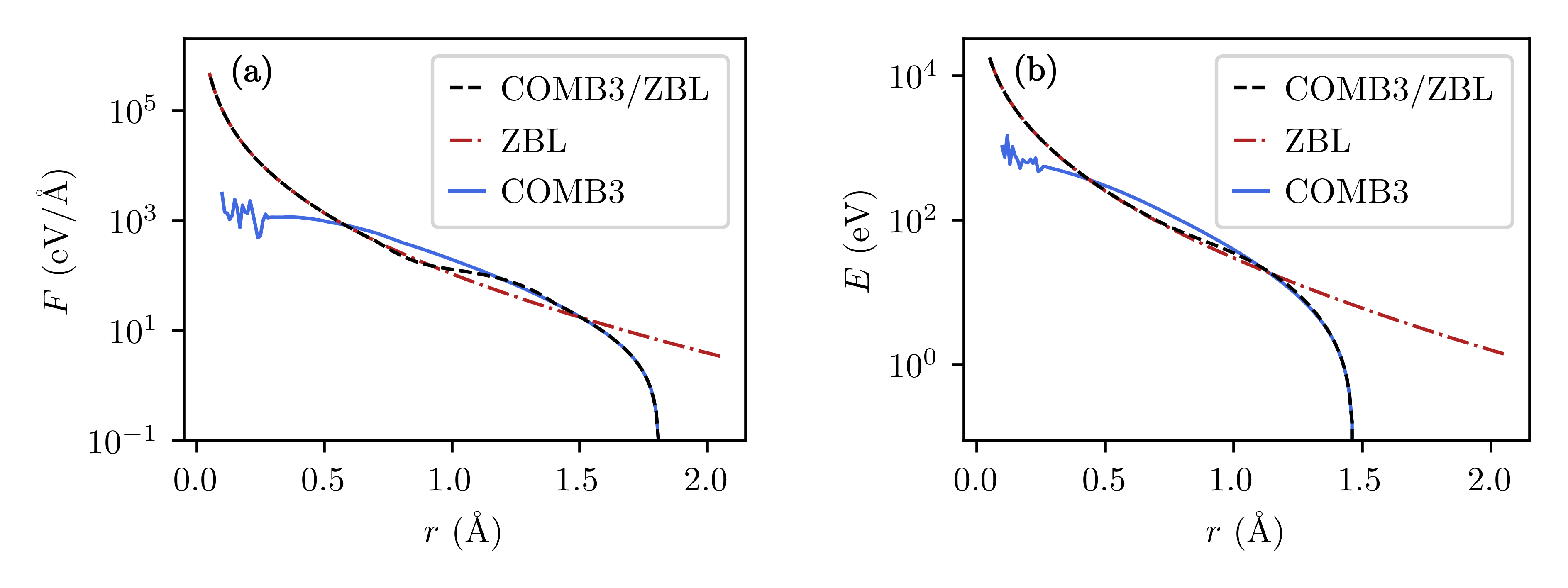}
\caption{\textbf{(a)} Force $F$ between and \textbf{(b)} potential energy of an Al adatom atop an Al wurtzite AlN(0001) surface atom. The contribution of the remaining surface is excluded. Both properties are presented as a function of the interatomic distance $r$.}
\label{fig:AlN} 
\end{figure}

Third, an Al atom on top the wurtzite AlN(0001) surface is lowered until it overlaps with a N surface atom. This simulation is performed twice. In the second run, interactions between the Al adatom and N surface atom are excluded. The difference of both runs are used to obtain the Al-N pair interaction for this many-body scenario. $F$ and $E$ are shown as a function of the interatomic distance in Fig.~\ref{fig:AlN} \textbf{(a)} and \textbf{(b)}, respectively.

\begin{figure}[ht]
\centering
\includegraphics[width=\linewidth]{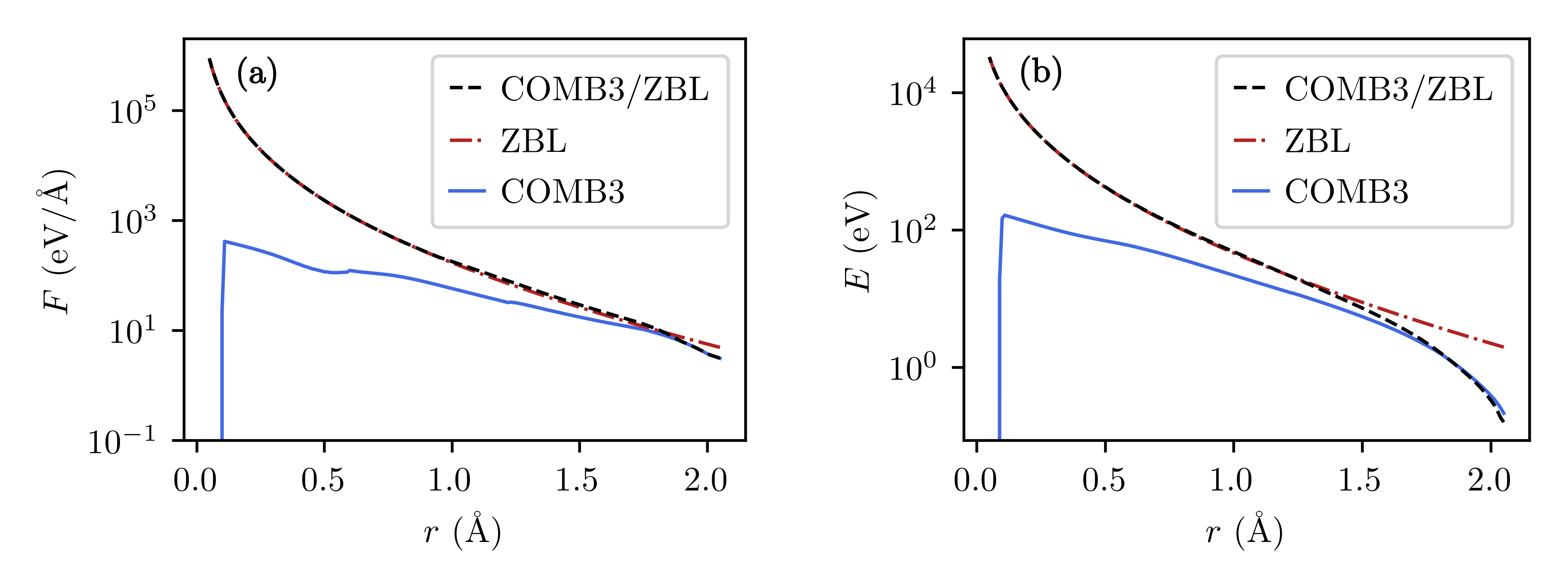}
\caption{\textbf{(a)} Force $F$ between and \textbf{(b)} potential energy of an Al adatom atop an Al wurtzite AlN(0001) surface atom. The contribution of the remaining surface is excluded. Both properties are presented as a function of the interatomic distance $r$.}
\label{fig:AlAl} 
\end{figure}

Fourth, an Al atom on top the wurtzite AlN(0001) surface is lowered until it overlaps with an Al surface atom. This simulation is performed twice. In the second run, interactions between the Al adatom and Al surface atom are excluded. The difference of both runs are used to obtain the Al-Al pair interaction for this many-body scenario. $F$ and $E$ are shown as a function of the interatomic distance in Fig.~\ref{fig:AlAl} \textbf{(a)} and \textbf{(b)}, respectively.

\clearpage
\section*{Supplementary Tables}

\begin{table}[ht] 
\centering
\begin{tabular}{|l|c|c|c|}
\hline
Al$_x$N$_y$ cluster structure & COMB3$_\mathrm{QTE^+}^{2022}$ & COMB3$_\mathrm{QEq}^{2016^*}$ & DFT/Exp. \\
\hline
N$_2$								& -4.90 & -4.90 & -4.88\cite{herzberg_molecular_1950}, -5.06\cite{nayak_evolution_1998} \\
\hline
Al$_1$N$_1$											& -1.67 & -4.37 & -1.38\cite{costales_density_2003}, -1.39\cite{nayak_evolution_1998}, -1.54\cite{herzberg_molecular_1950} \\
\hline
Al$_1$N$_2$	(Al-N-N,$C_s$\cite{kandalam_first_2000})										& -3.13 & -4.61 & -3.52\cite{kandalam_first_2000}\\
\hline
Al$_1$N$_2$	(N-Al-N,$D_{\infty h}$\cite{kandalam_first_2000})										& -1.53 & -4.74 & -1.80\cite{kandalam_first_2000}\\
\hline
Al$_1$N$_2$	(N-Al-N,$C_{2\nu}$\cite{kandalam_first_2000})										& -3.03 & -4.55 & -3.51\cite{kandalam_first_2000}\\
\hline
Al$_2$								& -0.64 & -0.64 & -0.77\cite{nayak_evolution_1998}, -0.78\cite{herzberg_molecular_1950} \\
\hline
Al$_2$N$_1$	(Al-N-Al,$C_{2\nu}$\cite{kandalam_first_2000})										& -2.51 & -3.43 & -2.59\cite{kandalam_first_2000}\\
\hline
Al$_2$N$_1$	(Al-Al-N,$C_{\infty\nu}$\cite{kandalam_first_2000})										& -1.65 & -3.16 & -1.41\cite{kandalam_first_2000}\\
\hline
Al$_2$N$_2$ (Is\cite{belbruno_structure_1999})		& -2.71 & -3.33 & -2.62\cite{belbruno_structure_1999} \\
\hline
Al$_2$N$_2$ (IIs\cite{belbruno_structure_1999})											& -2.27 & -3.78 & -3.08\cite{costales_chemical_2002}, -3.06\cite{belbruno_structure_1999} \\
\hline
Al$_2$N$_2$ (IIIs\cite{belbruno_structure_1999})	& -2.29 & -3.13 & -2.27\cite{belbruno_structure_1999} \\
\hline
Al$_2$N$_2$ (III\cite{costales_density_2003}$-$V\cite{costales_density_2003})		& -0.51 & -0.27 & 0.15\cite{costales_density_2003} \\
\hline
Al$_2$N$_2$ (IV\cite{costales_density_2003}$-$V\cite{costales_density_2003})		& 0.92 & -0.65 & 0.03\cite{costales_density_2003} \\
\hline
Al$_3$N$_2$ & -3.09 & -2.77 & -2.84\cite{costales_chemical_2002} \\
\hline
Al$_3$N$_2$ (b\cite{song_stable_2002}$-$a\cite{song_stable_2002}) & 0.45 & 0.01 & 0.00\cite{song_stable_2002} \\
\hline
Al$_3$N$_3$ (e\cite{kandalam_theoretical_2001}, VIIs\cite{belbruno_structure_1999}) & -2.90 & -4.31 & -3.28 \cite{kandalam_theoretical_2001}, -3.21\cite{belbruno_structure_1999} \\
\hline
Al$_3$N$_3$ (VIIIs\cite{belbruno_structure_1999})	& -2.77 & -4.15 & -2.47\cite{belbruno_structure_1999} \\
\hline
Al$_3$N$_3$ (C\cite{costales_density_2003}$-$A\cite{costales_density_2003}) & 0.0 & 0.22 & 0.24\cite{costales_density_2003} \\
\hline
Al$_4$N$_2$ & -2.63 & -2.67 & -3.22\cite{costales_chemical_2002} \\
\hline
Al$_4$N$_4$ (Xs\cite{belbruno_structure_1999}) & -2.50 & -4.16 & -3.52 \cite{belbruno_structure_1999} \\
\hline
Al$_4$N$_4$ (XIs\cite{belbruno_structure_1999}) & -3.83 & -3.07 & -3.35 \cite{belbruno_structure_1999} \\
\hline
Al$_5$N$_2$ & -2.69 & -1.95 & -3.01\cite{costales_chemical_2002} \\
\hline
Al$_6$N$_2$ (Al$_6$N$_2$\cite{costales_chemical_2002}) & -3.50 & -2.44 & -2.85\cite{costales_chemical_2002} \\
\hline
Al$_6$N$_2$ (Al$_6$N$_2$-2\cite{costales_chemical_2002}) & -3.07 & -2.41 & -2.36\cite{costales_chemical_2002} \\
\hline
\end{tabular}
\caption{\label{table: cluster_Appendix}Binding energies $E_\mathrm{bind}$ (eV/atom) of Al$_x$N$_y$ cluster ($x\leq6$, $y\leq4$) computed with COMB3$_\mathrm{QTE^+}^{2022}$ and COMB3$_\mathrm{QEq}^{2016^*}$ are compared with DFT and experimental findings. Labels in brackets allow for a structure identification in the provided reference when necessary. The subtraction of two labels indicates a relative binding energy.}
\end{table}

\begin{table}[ht]
\centering
\begin{tabular}{|l|c|c|c|c|}
\hline
Wurtzite & COMB3$_\mathrm{QTE^+}^{2022}$ & COMB3$_\mathrm{QEq}^{2016^*}$& COMB3$_\mathrm{QEq}^{2016}$  & DFT/Exp. \\
\hline
$E_\mathrm{f}^\mathrm{v_{Al}}$ (eV) 					& 3.63 & -1.48 & 0.3\cite{choudhary_dynamical_2016} & 2.36\cite{hung_ab_2004}, 3.76\cite{hung_ab_2004}, 4.46\cite{hung_ab_2004}, 5.6\cite{gorczyca_theory_1999} \\
\hline
$E_\mathrm{f}^\mathrm{Al_i^T}$ (eV) 					& 9.14 & 11.10 & & 15.83\cite{stampfl_theoretical_2002} \\
\hline
$E_\mathrm{f}^\mathrm{Al_i^O}$ (eV) 					& 9.75 & 9.41 & & 14.30\cite{stampfl_theoretical_2002} \\
\hline
$E_\mathrm{f}^\mathrm{v_{N}}$ (eV)	 					& 6.33 & 3.16 & 4.8\cite{choudhary_dynamical_2016} & 5.36\cite{hung_ab_2004}, 6.33\cite{hung_ab_2004}, 6.46\cite{hung_ab_2004}, 6.6\cite{gorczyca_theory_1999}\\
\hline
$E_\mathrm{f}^\mathrm{(\text{N-N})_N}$ (eV) 					& 5.39 & -3.51 & & 5.05\cite{szallas_characterization_2014} \\
\hline
$E_\mathrm{f}^\mathrm{N_i^T}$ (eV) 						& 5.75 & -0.67 & &  \\
\hline
$E_\mathrm{f}^\mathrm{N_i^O}$ (eV) 						& 9.28 & 8.37 &  & \\
\hline
$\gamma_\mathrm{\text{N-(0001)}}$ (eV/\r A$^2$)				& 0.398 & 0.142 & & 0.364\cite{holec_surface_2012} \\
\hline
$\gamma_\mathrm{\text{Al-(0001)}}$ (eV/\r A$^2$)			& 0.127 & 0.102 & & 0.136\cite{jindal_density_2009} \\
\hline
$\gamma_\mathrm{(1\overline{1}00)}$ (eV/\r A$^2$)	& 0.130 & 0.110 & 0.133\cite{choudhary_dynamical_2016} & 0.145\cite{holec_surface_2012} \\
\hline
$\gamma_\mathrm{(11\overline{2}0)}$ (eV/\r A$^2$)	& 0.156 & 0.117 & & 0.187\cite{holec_surface_2012} \\
\hline
\end{tabular}
\caption{\label{table: WurziteAppendix}Wurtzite AlN defect formation and surface energies obtained with COMB3$_\mathrm{QTE^+}^{2022}$ and COMB3$_\mathrm{QEq}^{2016^*}$ are compared to the COMB3$_\mathrm{QEq}^{2016}$ AlN publication, and  DFT as well as experimental findings.}
\end{table}

\begin{table}[ht]
\centering
\begin{tabular}{|l|c|c|c|c|}
\hline
Zinc blende & COMB3$_\mathrm{QTE^+}^{2022}$ & COMB3$_\mathrm{QEq}^{2016^*}$& DFT/Exp. \\
\hline
$E_\mathrm{f}^\mathrm{v_{Al}}$ (eV) 					& 3.32 & -1.74 & 2.28\cite{stampfl_theoretical_2002} \\
\hline
$E_\mathrm{f}^\mathrm{Al_N}$ (eV) 						& 11.25 & 7.53 & 15.92\cite{stampfl_theoretical_2002} \\
\hline
$E_\mathrm{f}^\mathrm{Al_i}$ (eV) 							& 6.01 & 8.59 & 7.69\cite{stampfl_theoretical_2002} \\
\hline
$E_\mathrm{f}^\mathrm{v_{N}}$ (eV)	 					& 5.95 & 2.91 & 5.73\cite{stampfl_theoretical_2002} \\
\hline
$E_\mathrm{f}^\mathrm{N_{Al}}$ (eV) 						& 5.37 & 5.91 & 5.37\cite{stampfl_theoretical_2002} \\
\hline
$E_\mathrm{f}^\mathrm{(\text{N-N})_N}$ (eV) 					& 5.22 & -4.24 & 5.37\cite{stampfl_theoretical_2002} \\
\hline
$E_\mathrm{f}^\mathrm{N_i}$ (eV) 						& 6.59 & 3.42 & 6.59\cite{stampfl_theoretical_2002} \\
\hline
$\gamma_\mathrm{\text{N-(100)}}$ (eV/\r A$^2$)				& 0.215 & 0.143 & 0.345\cite{holec_surface_2012} \\
\hline
$\gamma_\mathrm{(110)}$ (eV/\r A$^2$)				& 0.148 & 0.098 & 0.141\cite{holec_surface_2012} \\
\hline
$\gamma_\mathrm{(112)}$ (eV/\r A$^2$)				& 0.171 & 0.107 & 0.232\cite{holec_surface_2012} \\
\hline
\end{tabular}
\caption{\label{table: zinc blende_Appendix}Zinc blende AlN defect formation and surface energies obtained with COMB3$_\mathrm{QTE^+}^{2022}$ and COMB3$_\mathrm{QEq}^{2016^*}$ are compared to the COMB3$_\mathrm{QEq}^{2016}$ AlN publication, and  DFT as well as experimental findings.}
\end{table}

\begin{table}[ht]
\centering
\begin{tabular}{|l|c|c|c|c|}
\hline
Rock salt & COMB3$_\mathrm{QTE^+}^{2022}$ & COMB3$_\mathrm{QEq}^{2016^*}$ & DFT/Exp. \\
\hline
$\gamma_\mathrm{(100)}$ (eV/\r A$^2$)				& 0.130 & 0.170 &  0.083\cite{holec_surface_2012} \\
\hline
$\gamma_\mathrm{(110)}$ (eV/\r A$^2$)				& 0.188 & 0.166 &  0.194\cite{holec_surface_2012} \\
\hline
$\gamma_\mathrm{(112)}$ (eV/\r A$^2$)				& 0.171 & 0.097 &  0.270\cite{holec_surface_2012} \\
\hline
\end{tabular}
\caption{\label{table: rock salt Appendix}Rock salt AlN surface energies obtained with COMB3$_\mathrm{QTE^+}^{2022}$ and COMB3$_\mathrm{QEq}^{2016^*}$ are compared to the COMB3$_\mathrm{QEq}^{2016}$ AlN publication, DFT, and experimental references.}
\end{table}

\clearpage
\section*{Supplementary Results}

\begin{figure}[ht]
\centering
\includegraphics[width=\linewidth]{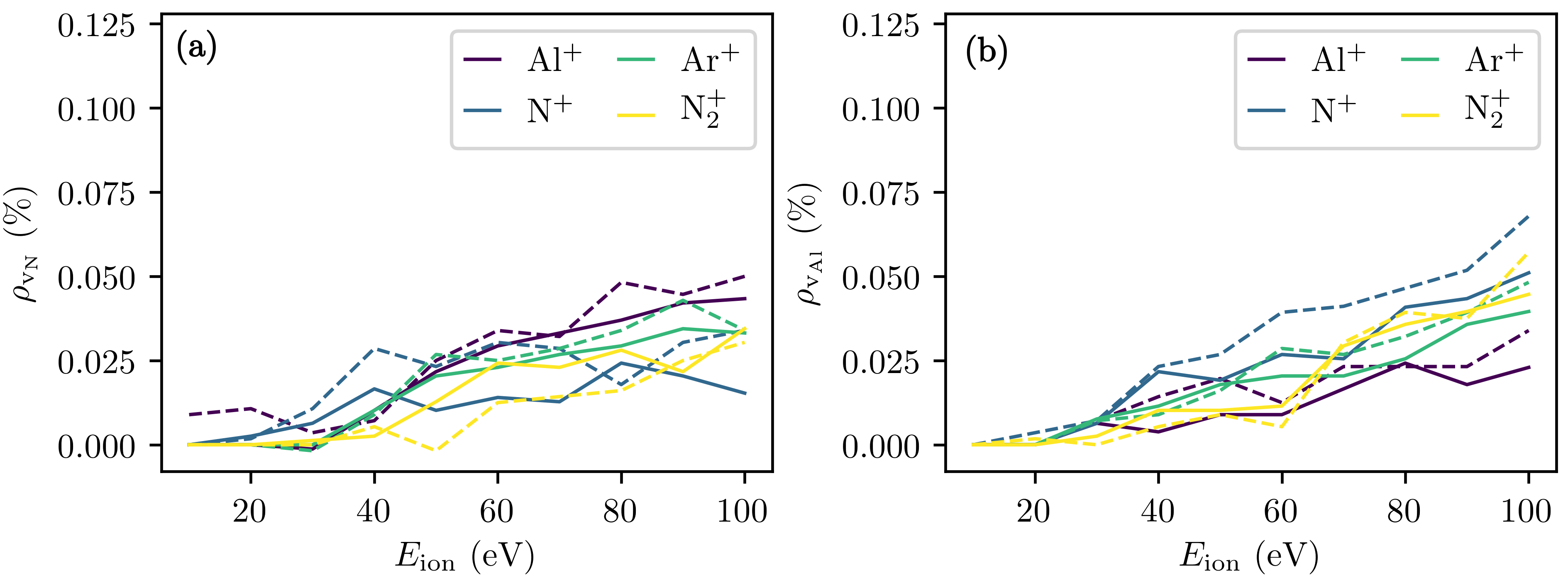}
\caption{\textbf{(a)} Nitrogen vacancy population $\rho_\mathrm{v_N}$ and \textbf{(b)} aluminum vacancy population $\rho_\mathrm{v_{Al}}$ is presented as a function of the ion energy $E$ and ion species (i.e., Al$^+$, N$^+$, Ar$^+$, N$_2^+$). Solid and dashed lines indicate surfaces initially equilibrated at 300 K and relaxed at 0 K, respectively.}
\label{fig:V} 
\end{figure}


\clearpage
\bibliography{./references.bib}
